\documentclass[12pt,a4paper]{article}

\usepackage[T1]{fontenc}
\usepackage[USenglish]{babel}   
\usepackage[margin=1in]{geometry}

\usepackage{graphicx}
\usepackage{braket}
\usepackage{xcolor}
\usepackage{subcaption}
\usepackage{multirow}

\usepackage{amsmath,amssymb,amsthm}
\usepackage{dsfont}

\usepackage{tikz}
\usetikzlibrary{calc}
\usetikzlibrary{hobby}
\usetikzlibrary{decorations.pathmorphing}

\usepackage{array}
\usepackage{hyperref}
\usepackage{etoolbox}
\makeatletter
\patchcmd{\BR@backref}{\newblock}{\newblock(}{}{}
\patchcmd{\BR@backref}{\par}{)\par}{}{}
\makeatother

\numberwithin{equation}{section}
\newtheoremstyle{break}{9pt}{9pt}{\itshape}{}{\bfseries}{}{\newline}{}
\theoremstyle{break}

\graphicspath{ {./plots/} }

\renewcommand{\Im}[0]{ \operatorname{Im} }

\begin{document}

\begin{flushleft}
{\bfseries\sffamily\Large 
Anchored random clusters and SLE excursions
\vspace{1.5cm} \\
\hrule height .6mm
}

\vspace{1.5cm}

{\bfseries\sffamily 
Federico Camia$^{1,2}$,
Valentino F.\ Foit$^{1,3}$,
Rongvoram Nivesvivat$^{1}$
}

\vspace{4mm}

{\small\textit{\noindent
$^1$ Science Division, New York University Abu Dhabi, Saadiyat Island, Abu Dhabi, United Arab Emirates \\
$^2$ Courant Institute of Mathematical Sciences, New York University, 251 Mercer Street, New York, NY 10012, USA \\
$^3$ Center for Cosmology and Particle Physics, New York University, 726 Broadway, New York, NY 10003, USA
}}

\vspace{4mm}

{\textit{E-mail:} \texttt{
federico.camia@nyu.edu,
foit@nyu.edu,
rongvoramnivesvivat@gmail.com
}}

\end{flushleft}
\vspace{7mm}

{\noindent\textsc{Abstract:}
We provide a pedagogical review of CFT techniques to compute certain Schramm-Loewner Evolution (SLE) observables in the upper half-plane. The approach relies on the ability to express the observables as bulk-boundary correlation functions that involve degenerate boundary operators and, therefore, obey certain differential equations. In particular, we recover Schramm's left-passage probability for SLE, the SLE Green's functions, and the generalized densities of ``anchored'' critical percolation clusters first obtained by Kleban, Simmons, and Ziff. We also obtain new formulas corresponding to the densities of pivotal points between critical Fortuin-Kasteleyn (FK) clusters.
}

\clearpage

\hrule 
\tableofcontents
\vspace{5mm}
\hrule
\vspace{5mm}

\section{Introduction}

In \cite{ksz06}, using conformal field theory methods, Kleban, Simmons, and Ziff compute the probability densities of critical percolation clusters in the upper half-plane, anchored to the real line. They do this by expressing the densities as correlation functions of local operators in conformal field theory (CFT), which obey consistency conditions such as positivity and boundary conditions. The latter can be determined by the geometry of the corresponding clusters while the former follows from the interpretation of the observables as, alternatively, densities or probabilities.

In this note, we revisit their method and extend it in a systematic way in several directions by considering
\begin{itemize}
    \item the densities of anchored clusters in the critical Fortuin-Kasteleyn (FK) random cluster model~\cite{fk72} with $Q\geq 1$, which reduces to critical percolation when $Q=1$,
    \item all non-vanishing bulk densities for clusters anchored to 2 marked points on the real axis,
    \item quantities corresponding to Schramm-Loewner Evolution (SLE) \cite{schramm99} observables that do not have an interpretation in terms of FK cluster densities. 
\end{itemize} 
The usefulness of the method stems from the fact that certain CFT correlation functions are solutions to the so-called BPZ equations \cite{bpz84}. The idea of expressing probabilities of connectivity events as correlation functions was used by Cardy in the derivation of his celebrated formula for the scaling limit of crossing probabilities in percolation \cite{Cardy92}.
Its validity is supported by the fact that it provides the correct answer in all cases where the answer is rigorously known, as in the case of Cardy's formula \cite{smirnov01}.

This note is meant to be pedagogical and is intended primarily as a review.
We will show how the use of the BPZ equations provides a unified derivation of a number of results that have appeared separately in the literature, including formulas that have been derived in the mathematics literature with completely different methods, involving the use of SLE techniques.
The plan of the note is summarized below:
\begin{itemize}
\item In Section \ref{sec:FK}, we briefly review the FK random cluster model, then we discuss the densities of anchored clusters that we wish to compute in Section~\ref{sec:anchored}. In Section~\ref{sec:normalization}, we explain our choice of normalization for the densities.
\item In Section \ref{sec:sle}, we provide an overview of the connections between FK clusters, loop models, SLE, and CFT. 
\item In Section \ref{sec:cft}, we review CFT techniques required for computing the densities of anchored clusters: this includes a discussion on the spectrum and correlation functions of a CFT.
\item In Section \ref{sec:conpro}, we discuss how to compute certain densities as CFT correlation functions, then we compare some of our results to known results in the mathematics literature.
\end{itemize}

We point out that not all the results derived here have been obtained rigorously, and some are new (to the best of our knowledge). It is an interesting challenge for \emph{SLExperts} and the mathematics community more generally to provide rigorous derivations.

\subsection{The FK random cluster model \label{sec:FK}}

The Fortuin-Kasteleyn (FK) random cluster model was introduced in \cite{fk72} as a generalization of the two-dimensional $Q$-state Potts model for non-integer $Q$.
For integer $Q$, the $Q$-state Potts model on a graph $\mathcal{L}$ is a spin model where the spin $\sigma(k)$, located at vertex $k \in \mathcal{L}$, takes values in $\{1, \ldots, Q\}$. The partition function of the $Q$-state Potts model is given by
\begin{equation} \label{zpotts1}
    Z_{\text{Potts}} = \sum_{\{\sigma(k)\}}e^{K\sum_{i,j}\delta_{\sigma(i),\sigma(j)}} \ ,
\end{equation}
where the sum in the exponent is taken over all the pairs of neighboring vertices, $\delta_{\cdot,\cdot}$ denotes the Kronecker delta, and $K$ is the coupling constant that determines the interaction between neighboring vertices.
From \cite{fk72}, the idea of generalizing \eqref{zpotts1} to non-integer $Q$ is to rewrite the sum over all possible spins in \eqref{zpotts1} as the sum over subgraphs $\mathcal{G}$ such that all the vertices in each maximal connected component have the same spin. This can be done by using the relation 
$e^{K\delta_{\sigma(i), \sigma(j)}} = 1+ (e^{K} - 1)\delta_{\sigma(i), \sigma(j)}$, which yields
\begin{equation} \label{zpotts2}
    Z_{\text{Potts}} = \sum_{\mathcal{G}}(e^{K} - 1)^{\#(\text{edges})}Q^{\#(\text{FK clusters})} \ ,
\end{equation}
where the \emph{FK clusters} are maximal connected components of $\mathcal{G}$ and $Q$ can now be interpreted as a cluster weight. The partition function \eqref{zpotts2} now makes sense for non-integer $Q$, in contrast to \eqref{zpotts1}, where the sum over all possible spins can only be defined for integer $Q$.

The FK random cluster model has become a prominent model in statistical physics due to its relation with the $Q$-state Potts model and bond percolation, corresponding to $Q=1$, and to its rich structure. The infinite-volume model exhibits a continuous phase transition at the critical coupling $K_c = \log( \sqrt{Q} + 1)$ for $0\leq Q\leq 4$. The critical model is scaling invariant and its continuum scaling limit is believed to enjoy full two-dimensional infinite conformal symmetry. While conformal symmetry has been rigorously established only for the Ising model on the square lattice \cite{smirnov07} (later extended to isoradial graphs \cite{chelkak2012universality}) and for site percolation on the triangular lattice \cite{smirnov01} (which is believed to be in the same universality class as the FK model with $Q=1$), assuming full conformal symmetry has a very powerful implication: it suggests that the scaling limit of the FK random cluster model converges to a conformally-invariant quantum field theory, known as conformal field theory (CFT). By assuming full conformal symmetry, theoretical physicists have been able to use methods from CFT to predict many of the model's physical observables, such as critical exponents and crossing probabilities, which so far appear to be in excellent agreement with the results obtained via rigorous approaches in mathematics. For instance, the formula for crossing probabilities in critical percolation obtained by Cardy~\cite{Cardy92} using CFT methods was later proved rigorously by Smirnov~\cite{smirnov01}. In this note, in the same spirit, we will use CFT techniques to compute the scaling limit of certain densities for anchored clusters in the FK random cluster model, to be discussed in Section \ref{sec:anchored}, some of which have been obtained by rigorous mathematical methods.
The fact that our derivation reproduces known mathematical results provides strong evidence for the validity of the CFT assumptions and approach, as well as for the correctness of the formulas that do not yet have a rigorous mathematical derivation.

\subsection{Anchored clusters} \label{sec:anchored}

We are interested in the continuum scaling limit of Bernoulli percolation and the critical FK random cluster model defined on a lattice in the upper half-plane, which, as explained above, should be described by a CFT.
We call \emph{anchored clusters} FK clusters in the upper half-plane that touch the real axis at prescribed locations. In particular, we will be interested in the probabilities of certain events associated with anchored clusters. In the continuum scaling limit, in order to keep them from vanishing, such probabilities need to be renormalized (rescaled by an appropriate power of the lattice spacing). We will use the term ``densities'' for the limits of such renormalized probabilities.
For example, by the density of an anchored cluster we mean the probability that a point of the upper half-plane belongs to the cluster, as a function of the location.
This type of density is studied in~\cite{ksz06} for the case of a percolation cluster pinned to a point or a segment on the real line (see also \cite{CamiaApr24} for a rigorous derivation of one one of the formulas in~\cite{ksz06}).

Here, we are interested in the same type of question, but we will consider more general densities, such as of cluster boundaries and pivotal points, for anchored clusters in the FK random cluster model. From a CFT's perspective, the densities associated to anchored clusters are encoded by correlation functions involving the insertion of a bulk field, where the density is computed, and of boundary fields which determine where and how the cluster is anchored to the real line.
The latter are degenerate fields whose properties will be discussed in Section~\ref{sec:cft-spec}. Correlation functions involving degenerate fields are solutions to the Belavin-Polyakov-Zamolodchikov (BPZ) equations of~\cite{bpz84}, which, in some special cases, reduce to the same ordinary differential equations encountered in the theory of $\text{SLE}$. Therefore, our final results will provide solutions to the BPZ equations that can be interpreted in terms of $\text{SLE}$ probabilities. 

In this note, we restrict ourselves to situations where the quantities of interest depend on the locations of three marked points, two on the real line and one in the upper half-plane. Consequently, the quantities we are interested in will be expressed as bulk-boundary three-point functions for certain fields that will be chosen appropriately.
The two marked points on the real line can represent either locations where a cluster is pinned to the real line or where an interface is inserted, corresponding to a change of boundary condition.
The point in the bulk corresponds to the location where the density under consideration is calculated, and we will consider various densities, as mentioned above and discussed in more detail below.

We will consider situations where one or more clusters are anchored to the real line in such a way that their boundaries produce $m$ lines at each of the two marked points on the real line, which will be assumed to be at locations $-\frac{L}{2}$ and $\frac{L}{2}$.
In the case of a single cluster, the ``density'' of the cluster at a bulk point $z$ will be denoted by $\rho_{m, m; 0}(L,z)$.
More generally, $\rho_{m, m; n}(L,z)$ will denote the renormalized probability that $\frac{n}{2}$ cluster boundaries meet at point $z$.
The densities $\rho_{m, m; n}$ can be expressed as CFT correlation functions on the upper half-plane $\mathbb{H}$ as
\begin{subequations}
\begin{align}
\rho_{m,m;n}(L, z) &= 
\Braket{ \phi_m \left(\frac L2\right)\phi_m \left(-\frac L2\right) \psi_n(z) }_\mathbb{H}
\label{rho3pt1} \\
\rho_{m,m;0}(L, z) &=  \Braket{ \phi_m \left(\frac L2\right)\phi_m \left(-\frac L2\right) \sigma(z) }_\mathbb{H} \, , \label{rho3pt2}
\end{align}
\label{rho3pt}
\end{subequations}
where $\phi_m(x)$ and $\psi_n(z)$ are local operators that insert $m$ interfaces at a point $x$ on the real line (the boundary of the upper half-plane) and $n$ interfaces at a point $z$ in the upper half-plane (the bulk), respectively, and $\sigma$ is the operator that inserts a cluster. The definitions of $\phi_m$, $\psi_n$, and $\sigma$ and their details will be discussed in Section \ref{sec:cft}.

Since the expressions represented by~\eqref{rho3pt1} only contain operators that insert interfaces, also called \emph{leg operators}, they can be understood as SLE observables, without reference to clusters, and make sense also for values of the SLE parameter $\kappa$ that do not correspond to any FK cluster model.
In contrast, the cluster-insertion or density operator cannot be expressed in terms of interfaces and consequently the expressions represented by~\eqref{rho3pt2} need to be interpreted as FK observables.

\begin{table}[t]
    \centering
    \begin{tabular}{|>{$}c<{$}|>{$}c<{$}|>{$}c<{$}|}
        \hline 
        \text{Cluster configuration} & \text{Density}& \text{Eq.} \\
        
        \hline\hline \rule{0pt}{4ex}
        
        \begin{tikzpicture}[x = 2.5 mm, y = 1.0 mm, baseline=2mm]
            \pgfmathsetseed{420}
            \draw[line width=1pt] (-5,0) -- (5,0);
            \draw[decorate, decoration={random steps,segment length=4pt, amplitude=3pt}] (-3,0) to[out angle=70, in angle=180-80, curve through = {(-3,6) (-2,8) (1,6)}] (3,0);
            \filldraw[red] (-3,0) circle (2pt) node[black, anchor=north]{$$};
            \filldraw[red] (3,0) circle (2pt) node[black, anchor=north]{$$};
            \filldraw[blue] (0,3) circle (2pt) node[black, anchor=south]{$$};
        \end{tikzpicture}
      & \rho_{1,1;0}(L, z) & \eqref{110} \\[1.5ex]

        \hline\hline

        \begin{tikzpicture}[x = 2.5 mm, y = 1.0 mm, baseline=2mm]
            \pgfmathsetseed{420}
            \draw[line width=1pt] (-5,0) -- (5,0);
            \draw[decorate, decoration={random steps,segment length=4pt, amplitude=3pt}] (-3,0) to[out angle=70, in angle=180-80, curve through = {(-3,6) (-2,8) (1,6)}] (3,0);
            \filldraw[red] (-3,0) circle (2pt) node[black, anchor=north]{$$};
            \filldraw[red] (3,0) circle (2pt) node[black, anchor=north]{$$};
            \filldraw[black] (-2,8) circle (2pt) node[black, anchor=south]{$$};
        \end{tikzpicture}
        & \rho_{1,1;2}(L, z) & \eqref{112} \\[1.5ex]

        \hline\hline

        \begin{tikzpicture}[x = 2.5 mm, y = 1.0 mm, baseline=2mm] % 2L spin
            \pgfmathsetseed{420}
            \draw[line width=1pt] (-5,0) -- (5,0);
            \draw[decorate, decoration={random steps,segment length=4pt, amplitude=3pt}] (-3,0) to[out angle=180-30, in angle=30, curve through = {(-3,6) (-2,8) (2,7)}] (3,0);
            \draw[decorate, decoration={random steps,segment length=4pt, amplitude=3pt}] (-3,0) to[out angle=30, in angle=180-30, curve through = {(-2,1) (1,2)}] (3,0);
            \filldraw[red] (-3,0) circle (2pt) node[black, anchor=north]{$$};
            \filldraw[red] (3,0) circle (2pt) node[black, anchor=north]{$$};
            \filldraw[blue] (-1,5.5) circle (2pt) node[black, anchor=south]{$$};
        \end{tikzpicture}
        & \rho_{2,2;0}(L, z) & \eqref{220} \\[1.5ex]

        \hline\hline

        \begin{tikzpicture}[x = 2.5 mm, y = 1.0 mm, baseline=2mm] % 2L spin
            \pgfmathsetseed{25939}
            \draw[line width=1pt] (-5,0) -- (5,0);
            \draw[decorate, decoration={random steps,segment length=4pt, amplitude=3pt}] (-3,0) to[out angle=180-30, in angle=30, curve through = {(-3,5) (-2,8) (2,7)}] (3,0);
            \draw[decorate, decoration={random steps,segment length=4pt, amplitude=3pt}] (-3,0) to[out angle=30, in angle=180-30, curve through = {(-2,1) (0,3) (1,2)}] (3,0);
            \filldraw[red] (-3,0) circle (2pt) node[black, anchor=north]{$$};
            \filldraw[red] (3,0) circle (2pt) node[black, anchor=north]{$$};
            \filldraw[black] (0,3) circle (2pt) node[black, anchor=south]{$$};
        \end{tikzpicture}
        & \rho_{2,2;2}(L, z) & \eqref{lowerportion} \\[1.5ex]

        \hline\hline

        \begin{tikzpicture}[x = 2.5 mm, y = 1.0 mm, baseline=2mm] % 2L 4L
            \pgfmathsetseed{25939}
            \draw[line width=1pt] (-5,0) -- (5,0);
            \draw[decorate, decoration={random steps,segment length=4pt, amplitude=3pt}] (-3,0) to[out angle=140, in angle=110, curve through = {(-3,6) }] (0,4);
            \draw[decorate, decoration={random steps,segment length=4pt, amplitude=3pt}] (-3,0) to[out angle=40, in angle=-130, curve through = {(-1,2)}] (0,4);
            \draw[decorate, decoration={random steps,segment length=4pt, amplitude=3pt}] (3,0) to[out angle=50, in angle=80, curve through = {(3,6)}] (0,4);
            \draw[decorate, decoration={random steps,segment length=4pt, amplitude=3pt}] (3,0) to[out angle=140, in angle=-60, curve through = {(1,2)}] (0,4);
            \filldraw[red] (-3,0) circle (2pt) node[black, anchor=north]{$$};
            \filldraw[red] (3,0) circle (2pt) node[black, anchor=north]{$$};
            \filldraw[black] (0,4) circle (2pt) node[black, anchor=east]{$$};
        \end{tikzpicture}
        & \rho_{2,2;4}(L, z) & \eqref{224} \\[1.5ex]

        \hline
    \end{tabular}
    \caption{Cluster configurations and densities for several events. In the diagrams, lines denote interfaces and red dots represent the marked points on the real line. The bulk points where densities are calculated are indicated by blue dots for the cluster-insertion operator and black dots for leg operators.}
    \label{cluster_configurations}
\end{table}

In Table \ref{cluster_configurations}, we list the relevant cluster and interface configurations and the corresponding three-point functions, which describe the associated densities, as well as the equations that display their analytic expressions.
We consider only the cases corresponding to $m=1,2$ and $n=0,2,4$.
For other values, the fractal dimensions become negative, which implies that the corresponding configurations are not seen (with probability one) in the continuum limit.
We note that $n \geq 6$ implies a percolation arm-event with 6 or more arms, whose critical exponent is strictly larger than 2, leading to a negative fractal dimension (see, e.g., \cite{PhysRevLett.83.1359}).
Analogously, $m \geq 3$ implies a percolation arm-event with 3 or more arms at the boundary (the real line), whose critical exponent is 1 or greater, leading to zero or negative fractal dimension.
The values of the critical exponents mentioned above were first rigorously proved for percolation \cite{10.1214/EJP.v3-32}, with the proof later extended to the FK-Ising model in \cite{10.1214/16-EJP3452} and to more general FK models in \cite{Copin2021}.

\subsection{Normalization of the densities} \label{sec:normalization}

To normalize the density $\rho_{m,m; n}(L, z)$, we assume that the bulk field $\psi_{n}(z)$ reduces to the boundary field $\phi_j(0)$ as $z\to 0$.
In general, $j$ may not be equal to $m$, and we will determine $\phi_j$ for each $\psi_n$ in Section \ref{sec:conpro} by considering the geometry of the corresponding SLE curves. In CFT terms, this assumption, combined with~\eqref{rho3pt}, implies that there is a constant $C$ such that, as $z \to 0$,
\begin{equation} \label{limrho}
    \Braket{ \phi_m(x_1)\phi_m(x_2) \psi_n(z) }_{\mathbb{H}}
    = C |z - \bar z|^{\Delta_{\phi_j} - 2\Delta_{\psi_n}}
    \Braket{ \phi_m(x_1)\phi_m(x_2) \phi_j(0) }_{\mathbb{H}}  + \ldots \ .
\end{equation}
Equation~\eqref{limrho} is called a boundary \emph{operator product expansion} (OPE); its right-hand side contains the three-point function
\begin{equation} \label{cmmn}
    \Braket{ \phi_{m}(x_1)\phi_{m}(x_2)\phi_{j}(x_3) }_{\mathbb{H}} =
    \frac{C_{m,m;j}}{(x_1 - x_2)^{2\Delta_{m} -\Delta_{j}}(x_2 -x_3)^{\Delta_j}(x_1 - x_3)^{\Delta_{j}}} \, ,
\end{equation}
where $C_{m,m;j}$ is a constant, called a \emph{structure constant} in the CFT jargon.
We choose to normalize the density $\rho_{m,m;n}$ in such a way that $C=1$ in~\eqref{limrho}.
In addition, we also normalize two-point functions in such a way that
\begin{equation}
\Braket{ \phi_{m}(x_1)\phi_{m}(x_2) }_{\mathbb{H}} = \frac{1}{(x_1 - x_2)^{2\Delta_{m}}} \, .
\end{equation}
With these choices, the structure constants $C_{m,m;j}$ coincide with the so-called OPE coefficients, which can be extracted from the boundary four-point function $\braket{ \prod_{i=1}^4\phi_m(x_i) }$, which in turn can be computed by solving the BPZ equations while taking into account the physical boundary conditions.
In Section \ref{sec:norm}, we will see that $C_{m,m;j}$ can be expressed in terms of the matrix elements of the crossing transformation of solutions of the BPZ equations.

\section{Loop models, SLE, and CFT {\label{sec:sle}}}

With the term \emph{loop models}, we refer to two-dimensional statistical mechanics models that can be represented as ensembles of non-intersecting loops. Notable examples are models of self-avoiding loops and the $O(n)$ loop model \cite{Nienhuis82}, as well as the collection of interfaces in percolation, the Ising model and the FK random cluster model. 
The loop model's picture allows us to apply well-developed SLE and CFT tools to the study of the FK random cluster model.

\begin{figure}[t]
    \centering
    \includegraphics[scale =0.5]{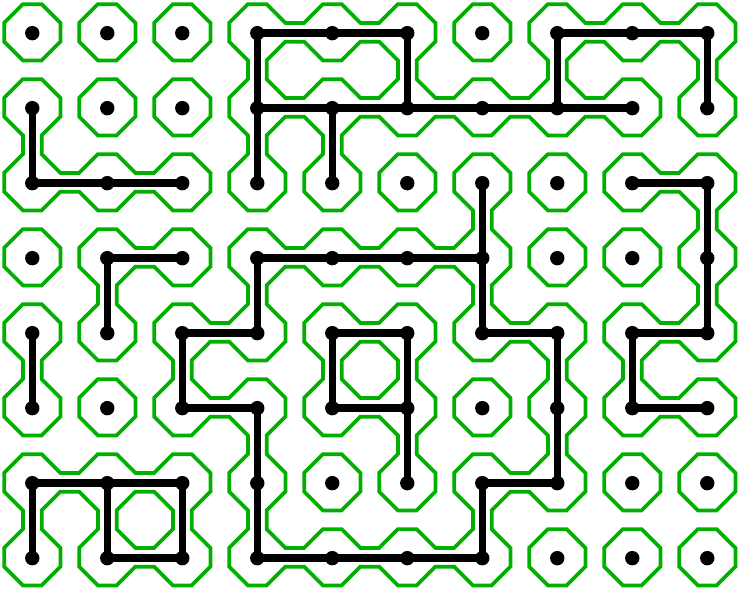}
    \caption{Example of FK clusters in black and their boundaries (loops) in green.}
    \label{fig:PottsLoops}
\end{figure}

It is shown in \cite{bkw76} that the FK random cluster model is in a one-to-one correspondence with the completely packed loop model on the square lattice, where the \emph{loops} are the interfaces between the connected subgraphs of $\mathcal{G}$ (the FK clusters) in \eqref{zpotts2} and the dual connected subgraphs in the dual lattice (the dual clusters), so that the loops can be considered boundaries of FK clusters (see Figure~\ref{fig:PottsLoops}). The partition function in \eqref{zpotts2} can be rewritten as a sum over such loops, defined on the medial lattice of the original square lattice \cite{bkw76}, and is given by
\begin{align} \label{zpotts3}
    Z_{\text{Potts}} \propto
    \sum_{\text{loops}}\left(\frac{e^K - 1}{n}\right)^{\#(\text{bonds})} n^{\#(\text{loops} )} \ ,
\end{align}
where
\begin{align} \label{nQ}
     n = \sqrt{Q} \ .
\end{align}
The parameter $n$ can be interpreted as the fugacity of the loops.
The latter correspond to the green loops of Figure~\ref{fig:PottsLoops}, while the number of bonds in~\eqref{zpotts3} is the same as the total number of diagonal segments in those loops. (Note that the bonds can be understood as edges of the medial lattice, but in Figure~\ref{fig:PottsLoops} those edges have been modified to make the figure easier to understand.)

The loop representation \eqref{zpotts3} allows us to make connection between the FK random cluster model and CFT. This can be seen as follows: the loops from the ensemble of interfaces can be given an orientation based on whether they wind around an FK cluster or a dual cluster. The FK random cluster model can then be mapped into the solid-on-solid (SOS) model by introducing a height function, defined on the union of all the vertices and dual vertices, which increases or decreases when an interface is crossed, depending on the orientation of the interface. In the scaling limit, such height functions are expected to converge to the (bosonic) free field modified by background charges that can be described by CFT \cite{fsz87, Nienhuis82}. 
For the mathematics literature on the SOS model, see the recent paper \cite{laslier2024tilted} and references therein.

\subsubsection{Schramm-Loewner Evolution (SLE)}

The Schramm-Loewner evolution, usually denoted by $\text{SLE}_{\kappa}$, is a stochastic process in two dimensions, introduced by Oded Schramm \cite{schramm99}, which generates a one-parameter family of planar random curves, characterized by the parameter $\kappa \geq 0$. Generally speaking, the scaling limit of critical interfaces in two-dimensional statistical mechanics models is conjectured to lead to $\text{SLE}_{\kappa}$ curves.
This was proved in the case of percolation \cite{smirnov01,cn06} and Ising interfaces \cite{CRMATH_2014__352_2_157_0} and is supported by numerical evidence \cite{PhysRevLett.88.130601} for the self-avoiding walk, which is conjectured to converge to $\text{SLE}_{8/3}$ \cite{lawler2002}. This convergence has led to several mathematically rigorous results concerning these statistical mechanics models, for example, the computation of percolation critical exponents \cite{10.4310/MRL.2001.v8.n6.a4} and crossing probabilities \cite{dubedat2006euler} and results on the scaling limit of Ising correlation functions \cite{zbMATH06446409} and the Ising magnetization field \cite{cgn15,cjn20}.

For critical loop models in the upper half-plane, in the scaling limit, the loops are also expected to be described by variants of $\text{SLE}_{\kappa}$ \cite{kager2004guide}, with a precise relation between the SLE parameter $\kappa$ and the loop fugacity $n$:
\begin{equation} \label{nk}
    n = -2 \cos\left(\frac{4\pi}{\kappa}\right) \quad \text{for} \quad 2\leq \kappa \leq 8 \ .
\end{equation}

The study of SLE has become a very active area of research in probability theory (see, e.g., \cite{lawler2008conformally, tsirelson2004lectures, Cardy_2005}). Here, we only mention a few facts that are directly relevant to our analysis.
The one-parameter family of $\text{SLE}_{\kappa}$ curves can be divided in two groups, depending on the value of $\kappa$,
\begin{align}
    \text{SLE}_{\kappa} \text{ generates}
    \begin{cases}
        \text{simple curves} &\text{for} \quad 0 \leq \kappa \leq 4 \\
        \text{non-simple curves} &\text{for} \quad \kappa > 4
    \end{cases}
\end{align}
Non-simple curves are curves that touch but do not cross themselves. For $\kappa \ge 8$, the curves become completely space-filling. From \eqref{nQ}, we see that $0 \leq n \leq 2$, since $Q$ takes values between $0$ and $4$. Therefore, \eqref{nk} tells us that the FK random cluster model is related to non-simple $\text{SLE}_{\kappa}$, with $\kappa>4$. The dilute phase of the $O(n)$ loop model \cite{Nienhuis82}, which includes the self-avoiding walk as a special limit, is related to simple $\text{SLE}_{\kappa}$, with $\kappa\leq 4$. The relations of the different phases of the $O(n)$ model to SLE are summarized in \eqref{domb}.

\begin{figure}[t]
    \centering
    \begin{subfigure}[t]{0.4\textwidth}
    \includegraphics[width=\textwidth]{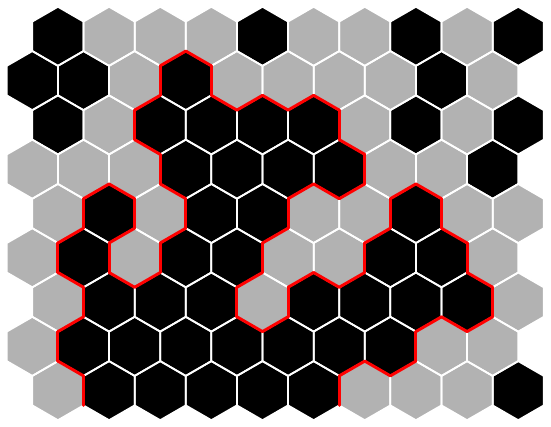}
        \caption{A percolation configuration with a cluster touching the real line and its boundary in the upper half-plane (in red).}
        \label{fig:fjord_perc}
    \end{subfigure} \hfill
    \begin{subfigure}[t]{0.4\textwidth}
        \includegraphics[width=\textwidth]{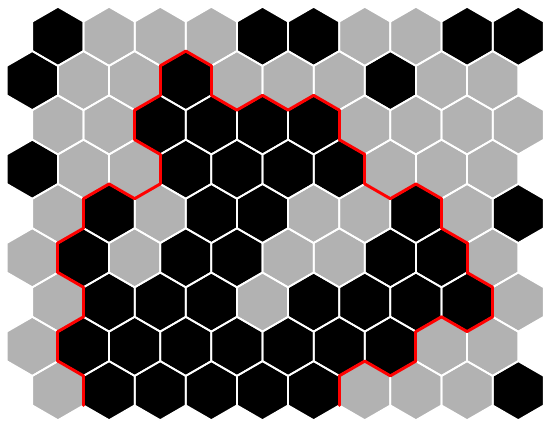}
        \caption{A percolation configuration with a cluster touching the real line. The external perimeter of the cluster (in red) corresponds to the boundary of its hull.}
        \label{fig:fjord_saw}
    \end{subfigure}
    \caption{A percolation cluster in the upper half-plane touching the real line (a) and its hull (b).}
    \label{fjords}
\end{figure}

It is also worth mentioning that there is a duality between the simple $\text{SLE}_{\kappa'}$ and the non-simple $\text{SLE}_{\kappa}$, provided that $\kappa \kappa' = 16$. Namely, the boundary of the hull of the non-simple $\text{SLE}_{\kappa}$ is described by a variant of the simple $\text{SLE}_{\kappa'}$ (see Figure~\ref{fjords}, where a percolation cluster in the upper half-plane and its hull are depicted).

The probabilities of certain $\text{SLE}_{\kappa}$ events can be expressed as solutions of the same differential equations obeyed by the correlation functions of certain boundary operators in CFT.
More generally, the probabilities of $\text{SLE}_\kappa$ events automatically satisfy all the axiomatic properties of CFT correlation functions involving  boundary operators. This suggests a deep connection between statistical mechanics, $\text{SLE}_{\kappa}$, and CFT, and makes it possible to analyze the probabilities of SLE events either through stochastic calculus or CFT techniques. This type of connection was first explored by Bauer and Bernard in~\cite{Bauer2002, Bauer2006}. 

\subsubsection{CFT approach}

Two-dimensional CFTs are (Euclidean) quantum field theories that are invariant under conformal transformations.
The action of a conformal transformation can be expressed in terms of differential operators that satisfy the Witt algebra.
The Witt algebra has a nontrivial central extension, characterized by its \emph{central charge} $c$, called the Virasoro algebra, which is the natural symmetry Lie algebra of a two-dimensional CFT.

In~\cite{Nienhuis82}, the $O(n)$ model is mapped onto a solid-on-solid (SOS) model, whose large-scale critical behavior is expected to be equivalent to that of a Gaussian model or a Coulomb gas (CG) \cite{Kadanoff_1978, nienhuis1984critical, fms97}.
Using this mapping and the CG formalism, Nienhuis \cite{Nienhuis82} was able to link the central charge $c$ of critical loop models to the loop fugacity $n$. This gives
\begin{equation} \label{cn}
    c = 13 - 6\beta^2 - 6\beta^{-2}
    \quad\text{and}\quad
    n= -2\cos \pi\beta^2 \ ,
\end{equation}
where the parameter $\beta^2$ is known as the Coulomb gas coupling. Another important application of the Coulomb gas formalism to critical loop models is the determination of the CFT {\it spectrum}: the collection of fields belonging to the theory, obtained in \cite{fsz87, Cardy_2006}, which will be summarized in Section~\ref{sec:cft-spec}.

Comparing \eqref{cn} with \eqref{nk}, we find the relation between the Coulomb gas coupling $\beta^2$ and the parameter $\kappa$ of $\text{SLE}_\kappa$:
\begin{equation} \label{kbe}
    \kappa  = \frac{4}{\beta^2} \ .
\end{equation}
Similarly to what happens in the case of $\text{SLE}_\kappa$ curves, which can be divided into two geometric regimes depending on whether $0 \leq \kappa \leq 4$ (simple curves) or $\kappa > 4$ (self-intersecting curves), the $O(n)$ loop model exhibits two distinct macroscopic critical regimes: the dilute phase and the dense phase.  These macroscopic behaviors are governed by the loop weight (or monomer fugacity) $x$ on the lattice \cite{Jacobsen2009}. For small values of $x$, typical loops are finite and the correlation functions decay exponentially. As $x$ reaches a critical threshold $x_c$, the system undergoes a second-order phase transition into the \textit{dilute phase}, where macroscopic loops appear and correlation functions transition to power-law decays \cite{Nienhuis82, DUPLANTIER1989229}. If the loop weight is further increased beyond the critical threshold ($x > x_c$), the loops proliferate and pack tightly across the lattice, pushing the system into the \textit{dense phase}. In this regime, the system remains critical and scale-invariant, but it flows to a different renormalization group fixed point, resulting in a distinct set of critical exponents governing the power laws.

These dilute and dense phases correspond directly to the valid parameter ranges of the Coulomb gas coupling $\beta^2$ that yield mathematically sensible statistical weights for the lattice models (i.e., loop fugacities $n \in [-2, 2]$) \cite{Nienhuis82}. They are given by
\begin{align}
\beta^2 =
\begin{cases}
    \frac1\pi\arccos\left(-\frac n2\right) &\in (0, 1]
    \quad\text{dilute phase}
    \Longleftrightarrow
    \text{SLE}_{\kappa \leq 4} \\
    2 - \frac1\pi\arccos\left(-\frac n2\right) &\in [1, 2)
    \quad\text{dense phase}
    \Longleftrightarrow
    \text{SLE}_{\kappa \geq 4}
\end{cases}
\label{domb}
\end{align}
Note that, in the borderline case $\beta=1$, corresponding to $\kappa=4$, the dilute and dense phases coincide.

The CFT approach pursued here is based on the observation that certain probabilities of events involving $\text{SLE}_\kappa$ curves and of more general events involving clusters can be expressed in terms of correlation functions containing boundary operators, which can sometimes be determined as solutions to differential equations that follow from conformal invariance, with appropriate boundary conditions, which need to be compatible with the behavior of $\text{SLE}_\kappa$ curves or FK clusters in specific limits. This observation allows us to use CFT techniques, applied to the computation of correlation functions, to compute certain (renormalized) probabilities for SLE, loop models, and the scaling limits of the FK random cluster model.
These CFT techniques are discussed in detail in the next section.

\section{\label{sec:cft}CFT techniques}

In this section, we review the necessary ingredients for a CFT computation of the densities in~\eqref{rho3pt}. In the context of CFT, such a computation involves four main steps:
\begin{itemize}
\item[1.] Identify the fields in the right-hand side of~\eqref{rho3pt}. This involves understanding the geometric interpretation of bulk and boundary primary fields belonging to the CFT spectrum in terms of the insertion of interfaces.
\item[2.] Compute correlation functions of the relevant primary fields as solutions to the BPZ equations.
\item[3.] Identify the appropriate solutions to the BPZ equations by applying additional constraints, such as positivity and boundary conditions.
\item[4.] Compute the coefficient $C_{m,m;j}$ in \eqref{cmmn} for each density by solving the BPZ equation of the corresponding boundary four-point function. This is needed to fix the normalization of the density.
\end{itemize}
In this section, we summarize the details of each step; examples of explicit calculations will be shown in Section~\ref{sec:conpro}.

\subsection{\label{sec:cft-spec}CFT Spectrum}
We separate the spectrum of the CFT that describes critical loop models on the upper half-plane into two groups: the bulk spectrum and the boundary spectrum.
In both cases, in the physics literature the spectrum was obtained by using the Coulomb gas formalism, under the assumption that the model's partition function converges, in the continuum scaling limit, to a path integral involving free bosonic fields with some special boundary conditions.
See~\cite{Cardy_2006, Cardy_2005} for the boundary spectrum for the free and wired boundary conditions, to be discussed below, 
and~\cite{fsz87} for the bulk spectrum.
Remarkably, many of the Coulomb gas predictions for the scaling dimensions of geometrical operators have been rigorously proven using SLE, for example, the rigorous derivation of the Hausdorff dimension of the SLE trace \cite{10.1214/07-AOP364}.

The fields that make up the spectrum correspond to highest-weight representations of the Virasoro algebra and are characterized by their conformal dimensions $\Delta_{(r,s)}$, which can be parametrized by the Kac indices $(r,s)$:
\begin{equation} \label{deltp}
    \Delta_{(r,s)} = P^2_{(r,s)} - P^2_{(1,1)}
    \quad\text{with}\quad
    P_{(r,s)} = \frac12(r\beta - s\beta^{-1}) \ ,
\end{equation}
where $\beta$ is the parameter that appears in~\eqref{cn} and $r$ is an integer. In the so-called minimal models of \cite{bpz84}, $s$ is also an integer. However, as we will see in \eqref{b3}, in critical loop models, the second index $s$ can take non-integer values, provided that the product $rs$ is an integer.

\subsubsection{Bulk spectrum}

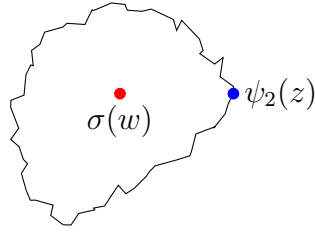
\begin{figure}[t]
\centering
\begin{tikzpicture}[x = 7.5 mm, y = 4.33 mm] % 2L 4L
    \pgfmathsetseed{123}
    \draw[decorate, decoration={random steps,segment length=4pt, amplitude=3pt}] (-3,0) to[out angle=140, in angle=110, curve through = {(-3,6) }] (0,4);
    \draw[decorate, decoration={random steps,segment length=4pt, amplitude=3pt}] (-3,0) to[out angle=40, in angle=-130, curve through = {(-1,2)}] (0,4);
    \filldraw[red] (-2,4) circle (2pt) node[black, anchor=north]{$\sigma(w)$};
    \filldraw[blue] (0,4) circle (2pt) node[black, anchor=west]{$\psi_2(z)$};
\end{tikzpicture}
\caption{The operator $\sigma(w)$ inserts a cluster at $w$, while the 2-leg operator $\psi_2(z)$ inserts 2 legs at $z$, which form a loop corresponding to the outer boundary of the cluster.}
\label{fplegs}
\end{figure}

To obtain the bulk spectrum, the authors of $\cite{fsz87}$ computed the torus partition functions of critical loop models, which include both the critical $O(n)$ loop model and the loop models associated with the critical FK cluster model. Their results give us the conformal dimensions for the bulk spectrum of critical loop models:
\begin{subequations}
\begin{align}
(\Delta_{(0, s)}, \Delta_{(0, s)})
\quad&\text{for}\quad s=k+\frac{1}{2} \text{ with } k \in \mathbb{N} 
\label{b1}
\\
(\Delta_{(1,s)}, \Delta_{(1,s)})
\quad&\text{for}\quad s \in \mathbb{N}^*
\\
(\Delta_{(r,s)}, \Delta_{(r, -s)})
\quad&\text{for}\quad r=k+\frac{1}{2} \text{ with } k \in \mathbb{N}^*
\quad\text{and}\quad s=\frac{j}{r} \text{ with } j \in \mathbb{Z} \ ,
\label{b3}
\end{align} 
\label{bspec}
\end{subequations}
where $\mathbb{N}^*$ denotes the set of positive integers.
Here we are only interested in the $2r$-leg operators, which we denote by $\psi_{2r}(z)$, where $z$ is a point in the upper half-plane and a \emph{leg} is a portion of a loop starting at $z$. For the loop models associated with the FK model, a leg is a portion of an interface that separates a cluster from surrounding dual clusters (i.e., a portion of a cluster boundary). For example, the operator $\psi_{2}(z)$ corresponds to inserting two legs at a point $z$ in the upper half-plane, as shown in Figure~\ref{fplegs}.
The $2r$-leg operators $\psi_{2r}(z)$ are primary fields whose conformal dimensions are given by~\eqref{b3} with $s=0$. More precisely, we have
\begin{equation}
    L_0 \psi_{2r}(z) = \bar L_0 \psi_{2r}(z) = \Delta_{(r,0)}\psi_{2r}(z) \ ,
\end{equation}
where $L_0$ is the dilatation generator in the Virasoro algebra of two-dimensional conformal transformations and the field $\psi_{2r}$ should be interpreted as a vector in a suitable representation of the algebra.

Another operator we are interested in is the spin operator $\sigma(z)$. The spin operator $\sigma(z)$ has left- and right-conformal dimensions given by~\eqref{b1} with $s=\frac12$:
\begin{equation} \label{sig}
    L_0 \sigma(z) = \bar L_0 \sigma(z) = \Delta_{(0, \frac12)}\sigma(z) \ .
\end{equation}
The operator $\sigma(z)$ corresponds to inserting a cluster at $z$ in the upper half-plane, therefore $\sigma$ is also known as the \emph{cluster-insertion operator}. For other conformal dimensions in \eqref{bspec}, the physical interpretation of the corresponding operators is not relevant to the context of this note, and we refrain from a detailed discussion, which can be found in \cite{gnjrs23}.

\subsubsection{Boundary spectrum}

In this note, we are interested in two types of boundary conditions in the FK random cluster model: \emph{free} and \emph{wired} boundary conditions. In terms of the $Q$-state Potts model, free boundary conditions correspond to having infinite temperature on the boundary.
In terms of the FK cluster model, this means that the boundary edges are all vacant. On the other hand, we can think of a system with wired boundary conditions as having zero temperature on the boundary: all vertices on the boundary belong to the same FK cluster and must have the same value of the Potts spin.

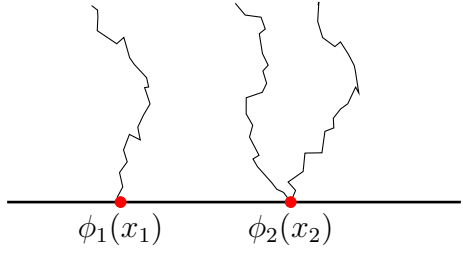
\begin{figure}[t]
\centering
\begin{tikzpicture}[x = 7.5 mm, y = 4.33 mm] % legs
    \pgfmathsetseed{420}
    \draw[line width=1pt] (-5,0) -- (3,0);
    \draw[decorate, decoration={random steps,segment length=4pt, amplitude=3pt}] (-3,0) to[out angle=120, in angle=-70, curve through = {(-2.5,3) }] (-3.5,6);
    \draw[decorate, decoration={random steps,segment length=4pt, amplitude=3pt}] (0,0) to[out angle=110, in angle=-20, curve through = {(-.7,2) (-0.5,4) }] (-1,6);
    \draw[decorate, decoration={random steps,segment length=4pt, amplitude=3pt}] (0,0) to[out angle=85, in angle=-110, curve through = {(1,3) (1,4.5) }] (0.5,6);
    \filldraw[red] (-3,0) circle (2pt) node[black, anchor=north]{$\phi_1(x_1)$};
    \filldraw[red] (0,0) circle (2pt) node[black, anchor=north]{$\phi_2(x_2)$};
\end{tikzpicture}
\caption{The boundary operator $\phi_1$ inserts 1 leg at point $x_1$ whereas the boundary operator $\phi_2$ inserts 2 legs at point $x_2$. In the context of the FK model, the legs represent interfaces (boundaries) between clusters.}
\label{blegs}
\end{figure}

From \cite{Cardy_2006, Cardy_2005}, the known boundary primary fields are the leg-insertion operators denoted by $\phi_{n}(x)$,
where we always use $x$ to denote a point on the real axis. The operators $\phi_{n}(x)$ have conformal dimensions $\Delta_{(n+1, 1)}$, where $n$ is a positive integer, that is to say,
\begin{equation} \label{dphi}
L_0 \phi_{n}(x) = \Delta_{(n+1, 1)}\phi_n(x)
\quad\text{for}\quad n \in \mathbb{N}^* \ .
\end{equation}
In contrast to the bulk case, there is only one copy of the Virasoro algebra for the boundary spectrum because the left- and right-moving Virasoro generators become identified on the real line due to the conformal boundary condition \cite{Cardy84}. Similarly to the $2r$-leg operators $\psi_{2r}$ in the bulk,
$\phi_{n}(x)$ corresponds to the insertion of $n$ legs at $x$ on the real line (see Figure~\ref{blegs}).

Furthermore, notice from \eqref{dphi} that the dimensions of the fields $\phi_n(x)$ are labeled by positive integers, that is, both $s=n+1$ and $r=1$ are positive integers. In unitary CFTs, primary fields whose dimensions are labeled by positive integers are \emph{degenerate}, meaning that they are annihilated by a particular combination $\mathcal{L}_{(s,r)}$ of Virasoro generators. The general expression for the null-vector operators $\mathcal{L}_{(r,s)}$ can be found in \cite{watts24}.
For instance,
\begin{subequations}
\begin{align}
\mathcal{L}_{(1,1)} & = L_{-1}
\\
\mathcal{L}_{(2, 1)} & = L_{-1}^2 -\beta^{2} L_{-2}
\\
\mathcal{L}_{(3, 1)} & = L_{-1}^3 -4\beta^{2} L_{-1}L_{-2} +4\beta^{2}\left(\beta^{2} + \frac12\right)L_{-3}
\ .
\end{align}
\label{exnull}
\end{subequations}

In non-unitary CFTs, such as those corresponding to critical loop models, fields with degenerate dimensions are not necessarily degenerate fields. Nevertheless, we will assume that the fields $\phi_n(x)$ are degenerate, i.e., we will assume that
\begin{equation} \label{vnull}
    \mathcal{L}_{(n+1, 1)}\phi_n(x)= 0 \ .
\end{equation}
A field $\phi_n$ satisfying \eqref{vnull} is called a \emph{level-$(n+1)$ degenerate field}. The fields $\mathcal{L}_{(n+1,1)}\phi_n$ are called the \emph{descendants} of $\phi_n$ and $\mathcal{L}_{(n+1, 1)}\phi_n$ is referred to as a \emph{null descendant}.
From \eqref{vnull}, the 0-leg operator $\phi_0$ can be identified with the identity operator, since $L_{-1}\phi_0(x)=0$.

A motivation for assumption \eqref{vnull} is the decoupling of the null descendants $ \mathcal{L}_{(n+1, 1)}\phi_n(x)$ from the annulus partition function \cite{Cardy_2006}. Moreover, the assumption that $\phi_n$ is a degenerate field is also strongly supported by successful comparisons between CFT and SLE computations --- see, for instance, the review article \cite{Cardy_2005}.

It is also worth mentioning that \eqref{vnull} strongly constrains the fusion rules between $\phi_m$ and $\phi_n$ \cite{bpz84}, leading to
\begin{equation} \label{degf}
    \phi_m \times \phi_n = \sum_{p=0}^{\min(m,n)} \phi_{|m-n|+2p} \ 
\end{equation}
where the fields are parametrized in terms of the number of legs inserted instead of the Kac indices in the standard convention. 

For generic Coulomb gas coupling $\beta^2$, it is also possible to understand \eqref{degf} from the point of view of $\text{SLE}_\kappa$. For instance, consider the \emph{product} (or \emph{fusion}, in the sense of an OPE) of two 1-leg operators $\phi_1$
\begin{equation}
    \phi_1(x_1) \times \phi_1(x_2) \stackrel{x_2 \to x_1}{\longrightarrow} \phi_0(x_1) + \phi_2(x_1) \ .
\end{equation}
This can be interpreted as follows. The two legs can coalesce near $x_1$, generating an excursion that vanishes as $x_2 \to x_1$, giving the identity operator $\phi_0(x_1)$, or they can coalesce away from $x_1$, generating a macroscopic loop pinned at $x_1$, resulting in the insertion of two legs at $x_1$, corresponding to $\phi_2(x_1)$.

For rational values of $\beta^2$, the fusion rules in \eqref{degf} may not hold, since the degenerate fields $\phi_n$ may acquire additional null descendants, due to the coincidence of Kac indices, and those null descendants may not vanish. For example, the identity $\phi_0$ has an extra null descendant at level 2 for $\beta^2 = \frac23$ (or $\kappa =6$). The emergence of non-vanishing null descendants usually results in more complicated fusion rules, which may include logarithmic representations of the Virasoro algebra, as studied in \cite{mr07}. 

This phenomenon arises naturally in two-dimensional critical percolation, which corresponds to $\kappa = 6$. In this case, the fusion rules of certain fields yield logarithmic singularities in addition to power laws, as shown in~\cite{cf24, cf24+}. This behavior reflects the indecomposable, non-diagonalizable nature of the Virasoro representations in logarithmic conformal field theories, which is mathematically necessary to correctly capture the full fusion algebra of percolation observables, as predicted in~\cite{GURARIE1993535, Vasseur_2012}.

It will prove useful later to parameterize the conformal dimensions \eqref{deltp} of the spin, bulk, and boundary $\ell$-leg operators in terms of $\kappa$ as
\begin{align}
\begin{alignedat}{2}
    [\sigma] &=    \Delta_{(0, \frac{1}{2})} &&= \frac{1}{2}-\frac{1}{\kappa}-\frac{3 \kappa}{64} \\
    [\psi_\ell] &= \Delta_{(\ell/2, 0)} &&= \frac{ 4 \ell^2 - (\kappa -4)^2 }{16 \kappa} \\
    [\phi_\ell] &= \Delta_{(\ell+1, 1)} &&= \frac{\ell(\ell+2)}{\kappa} - \frac{\ell}{2} \ .
\end{alignedat}
\end{align}

\subsection{Correlation functions on the upper half-plane}

In this section, we discuss how to compute CFT correlation functions on the upper half-plane, and we only focus on the correlation functions with at least 1 degenerate field evaluated at a point on the real axis, since they are relevant to study anchored clusters. In general, they take the form
\begin{equation} \label{hcorr}
    \Braket{ \phi_{n}(x) \prod_{j=1}^{b}\mathcal{O}_j(z_j) }_{\mathbb{H}} \ ,
\end{equation}
where the $\mathcal{O}_j$'s denote primary fields.

\subsubsection{From the full plane to the upper half-plane}

In any  two-dimensional CFT, correlation functions are solutions to the Ward identities: a system of linear differential equations that are consequences of conformal invariance. 
From the seminal paper \cite{Cardy84}, correlation functions of the form~\eqref{hcorr} can be computed using a calculation on the Riemann sphere $\mathbb{S}$ as input, where the Riemann sphere is defined as the complex plane $\mathbb{C}$ plus one point at infinity:
\begin{equation}
    \mathbb{S} = \mathbb{C}\cup \{\infty\} \ .
\end{equation}
In CFT on the upper half-plane, the correlation functions of the form~\eqref{hcorr} obey the same Ward identities as the following correlation functions on $\mathbb{S}$
\begin{equation} \label{fcorr}
    \Braket{ \tilde\phi_n(x) \prod_{j=1}^{b}\mathcal{O}_j(z_j)\mathcal{O}_j(\bar z_j) }_{\mathbb{S}} \ ,
\end{equation}
where $\tilde\phi_{n}(x)$ is the bulk-degenerate field located on the real axis in the full plane $\mathbb{S}$, whose left- and right-dimensions are identically $\Delta_{(n+1, 1)}$. Therefore, this allows us to write some correlation functions on the upper half-plane in terms of correlation functions on the full plane. For example, from  \cite{ksz06}, we have
\begin{equation} \label{23bh}
    \Braket{ \phi_1(x) \sigma (z) }_{\mathbb{H}} \propto \Braket{ \tilde\phi_1(x) \sigma (z)\sigma (\bar z) }_{\mathbb{S}} \ .
\end{equation}
However, we stress here that upper half-plane correlation functions and full-plane correlation functions do not always coincide as in \eqref{23bh} because, in addition to the Ward identities, correlation functions in CFT also obey other consistency conditions such as crossing symmetry and boundary conditions, and these addition constraints act differently on the full plane and the upper half-plane (see, for example, the case of the two-point function in Liouville theory on the disk and its corresponding four-point function in \cite{fzz00}).

\subsection{The BPZ equations}

 Equation~\eqref{vnull} implies that the correlation functions involving the degenerate field $\phi_{n}(x)$ satisfy certain differential equations, known as the BPZ equations \cite{bpz84}. In particular, for the correlation functions in \eqref{hcorr}, we have
\begin{equation}
\Braket{
\mathcal{L}_{(n+1, 1)}
\phi_{n}(x)
\prod_{j=1}^{b}\mathcal{O}_j(z_j)
}_{\mathbb{H}} 
=
\mathcal{D}_{(n+1, 1)}
\Braket{
\phi_{n}(x)
\prod_{j=1}^{b}\mathcal{O}_j(z_j)
}_{\mathbb{H}} 
 = 0
 \ ,
 \label{bch}
\end{equation}
where $\mathcal{L}_{(n+1, 1)}$ is a combination of Virasoro generators acting on fields, as in~\eqref{exnull}, and $\mathcal{D}_{(n+1, 1)}$ denotes a differential operator with respect to $x$, $z_j$, acting on correlation functions. Using the Ward identities \cite{fms97}, the expression of $\mathcal{D}_{(n+1, 1)}$ can be obtained by representing the Virasoro generators $L_{n}$ contained in $\mathcal{L}_{(n+1, 1)}$ as differential operators. 

Since the two correlation functions in \eqref{hcorr} and \eqref{fcorr} satisfy the same Ward identities, it is possible to show that both of them satisfy the same BPZ equations \cite{Cardy84}. Therefore, we have
\begin{equation} \label{fbh}
\mathcal{D}_{(n+1, 1)}
\Braket{
\tilde\phi_{n}(x)
\prod_{j=1}^{b}\mathcal{O}_j(z_j)\mathcal{O}_j(\bar z_j)
}_{\mathbb{S}}
= 0 \ ,
\end{equation}
where $\mathcal{D}_{(n+1, 1)}$ is the same differential operator as in \eqref{bch}. In practice, it is more convenient to write down $\mathcal{D}_{(n+1, 1)}$ from the Ward identities of the correlation functions \eqref{fcorr} on $\mathbb{S}$, due to the absence of a boundary. 

We will now review the second- and third-order BPZ equations \eqref{fbh} for the case of four-point functions that are relevant to the study of anchored clusters.

\subsubsection{The second-order BPZ equation}

We consider three-point functions of two 1-leg boundary operators and one bulk primary field on the upper half-plane. This type of three-point function corresponds to the densities in \eqref{rho3pt} with $m=1$.  Global conformal invariance dictates that such three-point functions take the form
\begin{equation} \label{11vv}
    \Braket{ \phi_1(x_1)\phi_1(x_2)\mathcal{O}(z) }_{\mathbb{H}}
    = \frac{C_{1,1;j}}{ (x_1 - x_2)^{2\Delta_{(2,1)}}(z -\bar z)^{2\Delta_{\mathcal{O}}}} G(\xi) \ ,
\end{equation}
where we have assumed that $\mathcal{O}(z) \to \phi_j(x)$ as $z \to x \in {\mathbb R}$. The function $G$ is not determined by global conformal invariance and is a function of the cross-ratio
\begin{equation} \label{xi}
    \xi = \frac{(x_1 - x_2)(z - \bar z)}{(x_1 - z)(x_2 - \bar z)} \ .
\end{equation}

Since $\phi_1(x_1)$ is a level-2 degenerate field, the three-point function \eqref{11vv} must obey a second-order BPZ equation. Following the discussion in the last subsection, we know that \eqref{11vv} obeys the same BPZ equation as the following bulk four-point function on the Riemann sphere:
\begin{equation} \label{bpz2}
    \mathcal{D}_{(2, 1)}
    \Braket{ \tilde\phi_1 (x_1) \tilde\phi_1 (x_2) \mathcal{O}(z)\mathcal{O}(\bar z) }_{\mathbb{S}} =  0 \ .
\end{equation}
As previously mentioned, the differential operator $\mathcal{D}_{(2,1)}$ can be obtained using the Ward identities, and we refrain from re-deriving it here because its expression is well-known and can be found in the literature (e.g.\ \cite{fms97}). Using \eqref{11vv}, we can then reduce \eqref{bpz2} to a second-order ordinary differential equation for the function $G(\xi)$, which leads to
\begin{equation}
\frac{d^2}{d \xi^2}
G(\xi)
+ \left\{
\frac{2+4\Delta_{(2,1)}(\xi - 2) -4\xi}{3\xi(1-\xi)}
\right\}
\frac{d}{d \xi}
G(\xi)
-\frac{(4\Delta_{(2,1)} + 2)\Delta_{\mathcal{O}}}{3(1-\xi)^2}
G (\xi) = 0 \ .
\label{bpz2nd}
\end{equation}

The above differential equation is an example of a second-order BPZ equation \cite{bpz84}. We denote $g_1(\xi)$ and $g_2(\xi)$ the two fundamental solutions, that is to say, a general solution $G(\xi)$ can be written as
\begin{equation} \label{2bpz}
    G(\xi) = c_1 g_1(\xi) + c_2g_2(\xi) \ .
\end{equation} 
The fundamental solutions to \eqref{bpz2nd} are well-studied in the literature and are the hypergeometric functions ${}_2F_1$.
Now, observe that \eqref{2bpz} is a differential equation with 3 singular points, namely at $\xi \in\{ 0, 1, \infty \}$. Consequently, there are 3 types of fundamental sets of solutions to \eqref{bpz2nd}, with different singular points. We will mainly be interested in solutions that are singular at $\xi = 0$ and whose asymptotic behaviors are given by
\begin{subequations}
\begin{align}
    g_1(\xi) &\overset{\xi \rightarrow 0}{\propto} 1 + \ldots \ , \\
    g_2(\xi) &\overset{\xi \rightarrow 0}{\propto} \xi^{\Delta_{(3,1)}} + \ldots \ .
\end{align}
\label{asymp}
\end{subequations}
The above asymptotic behaviors will be crucial for determining the boundary condition of the densities in \eqref{rho3pt}.

It is also possible to transform the set of solutions in \eqref{asymp} to other fundamental sets of solutions with different singular points.
The 3 types of fundamental sets of solutions are related by the so-called crossing-transformation. For instance, adding superscripts to $g_1$ and $g_2$ to indicate their singular points, we have
\begin{subequations}
\begin{align}
    g^{(0)}_1(\xi) &= F_{11}g^{(1)}_{1}(\xi) + F_{12}g^{(1)}_{2}(\xi) \\ 
    g^{(0)}_2(\xi) &= F_{21}g^{(1)}_{1}(\xi) + F_{22}g^{(1)}_{2}(\xi) \ ,
\end{align}
\label{F2b}
\end{subequations}
where the coefficients $F_{ij}$ are well-known \cite{fms97}. We only display their expressions for the special case $\Delta_{\mathcal{O}} =\Delta_{(2,1)}$, which will be relevant for the computation of the coefficient $C_{1, 1; 2}$ in Section~\ref{sec:norm}: 
\begin{align}
\left. \begin{pmatrix} F_{11} & F_{12} \\ F_{21} & F_{22} \end{pmatrix} \right|_{\Delta_{\mathcal{O}} = \Delta_{(2,1)}} =
\begin{pmatrix}
    -\frac{1}{2} \sec \left(\pi  \beta ^2\right) &
    \frac{\Gamma \left(1-2 \beta ^2\right) \Gamma \left(2-2 \beta ^2\right)}{\Gamma \left(2-3\beta^2\right) \Gamma \left(1-\beta ^2\right)} \\
    \frac{\Gamma \left(2 \beta ^2\right) \Gamma \left(2 \beta ^2-1\right)}{\Gamma \left(\beta ^2\right) \Gamma \left(3 \beta^2-1\right)} &
    \frac{1}{2} \sec \left(\pi  \beta ^2\right)
\end{pmatrix} \ .
\end{align}

\subsubsection{The third-order BPZ equation}

We now move to the three-point functions that describe the densities in \eqref{rho3pt} with $m=2$. This is a family of three-point functions involving two 2-leg operators and one bulk field, which can be written as follows:
\begin{align} \label{22vv}
    \Braket{ \phi_2(x_1) \phi_2(x_2) \mathcal{O}(z) }_{\mathbb{H}}
    = \frac{C_{2,2;j}}{(x_1 - x_2)^{2\Delta_{(3,1)}}(z -\bar z)^{2\Delta_{\mathcal{O}}}} H(\xi) \ ,
\end{align}
where we have assumed that $\mathcal{O}(z) \to \phi_j(x)$ as $z \to x \in \mathbb{R}$. Similarly to \eqref{11vv}, global conformal invariance determines the form of \eqref{22vv} up to an undetermined function $H$ of the cross-ratio $\xi$ defined in \eqref{xi}.

Following the same logic as in  the case of the second-order BPZ equations, the presence of the level-3 degenerate field $\phi_2(x_1)$ leads to a third-order BPZ equation for the function $H(\xi)$, namely
\begin{align} \label{bpz3rd}
\alpha(\xi)\frac{d^3}{d \xi^3}
H(\xi)
+ \beta(\xi)\frac{d^2}{d \xi^2} H(\xi)
+ \gamma(\xi)\frac{d}{d \xi} H(\xi)
+ \delta(\xi) H(\xi) = 0 \ ,
\end{align}
with
\begin{subequations}
\begin{align}
\alpha(\xi) &= \xi^2(\xi - 1)^2 \\
\beta(\xi) &= 2\xi(\xi-1)\left [ 2\xi - 1 - \Delta_{(3,1)}(\xi-2) \right] \\
\delta(\xi) &= 2\Delta_{\mathcal{O}}\Delta_{(3,1)} \left( \Delta_{(3,1)} + 1 \right)\frac{\xi(\xi-2)}{\xi-1}
\end{align}
and
\begin{align}
\begin{split}
\gamma(\xi) = 
& 3\Delta_{(3,1)}(\Delta_{(3,1)} - 1)
-(3\Delta_{(3,1)} - 1)(\Delta_{(3,1)} - 2)\xi \\
&+ \left[(\Delta_{(3,1)} -1)(\Delta_{(3,1)} -2) - 2\Delta_{\mathcal{O}}(\Delta_{(3,1)} + 1)\right]\xi^2 \ .
\end{split}
\end{align}
\end{subequations}

In contrast to the case of \eqref{bpz2nd}, whose fundamental solutions are well studied, closed-form expressions for the fundamental solutions to \eqref{bpz3rd} are not known in general. However, it is possible to express these solutions in terms of Coulomb-gas integrals through the Coulomb-gas formalism of Dotsenko and Fateev \cite{DF84}. In addition, Zamolodchikov in~\cite{zam84} worked out a recursion that provides expressions for generic Virasoro conformal blocks in terms of power series of the cross-ratio $\xi$, including the solutions to \eqref{bpz3rd} as special cases.

As in the case of the second-order BPZ equation discussed above, the third-order equation \eqref{bpz3rd} comes with 3 types of fundamental sets of solutions, corresponding to the 3 singular points at $\xi \in \{0, 1, \infty\}$. The crossing transformation between those 3 types of solutions is also known \cite{nr21} and will be crucial in the computation of the coefficients $C_{2,2; 2}$ and $C_{2,2; 4}$ in Section \ref{sec:norm}.

We will be mainly interested in solutions to \eqref{bpz3rd} with singular point at $\xi = 0$, which will be denoted by $h_1(\xi)$, $h_2(\xi)$, and $h_3(\xi)$. Therefore, we write a general solution $H$ as
\begin{align}
H(\xi) = a_1h_1(\xi) + a_2h_2(\xi) + a_3h_3(\xi) \ .
\end{align}
Using the degenerate fusion rules between $ \phi_2(x_1)$ and $ \phi_2(x_2)$ in \eqref{degf}, we can deduce that, as $\xi \rightarrow 0$, the above solutions behave as follows:
\begin{subequations}
\begin{align}
h_1(\xi ) &\overset{\xi \rightarrow 0}\propto 1 + \ldots \\
h_2(\xi ) &\overset{\xi \rightarrow 0}\propto \xi^{\Delta_{(3,1)}} + \ldots \\
h_3(\xi ) &\overset{\xi \rightarrow 0}\propto \xi^{\Delta_{(5,1)}} + \ldots \ .
\end{align}
\label{h123}
\end{subequations}
In the next section, we will provide closed and simple expressions for the solutions to \eqref{bpz3rd} for the primary fields that are relevant to the study of the densities in \eqref{rho3pt}.

\section{\label{sec:conpro} Densities of SLE excursions}

In this section, we compute  the densities $\rho_{m,m; n}(L, z)$ of \eqref{rho3pt}. Let us start by introducing some notation. Unless otherwise specified, from now on, the parameters $x_1$ and $x_2$ will always take the values
\begin{align}
    x_1 =  -\frac{L}{2} \quad \text{and} \quad x_2 =  \frac{L}{2} \, ,
\label{12z}
\end{align}
where $L$ is a positive number. The cross-ratio $\xi$ \eqref{xi} with \eqref{12z} becomes
\begin{equation}
\xi = \frac{4L(z -\bar z)}{(L -2 z)(L +2\bar z)} \ .
\label{12xi}
\end{equation}
Note that $\xi \rightarrow 0$ as $z \rightarrow \bar z$, and we will be using this limit to analyze the boundary conditions of the densities \eqref{rho3pt}.

\subsubsection{Physical solutions}

We identify the solutions to the BPZ equations that correspond to the densities $\rho_{m,m;p}(L, z)$ in \eqref{rho3pt} by requiring that they satisfy the following conditions:
\begin{itemize}
    \item They must be consistent with the properties of SLE curves: expanding the solution in powers of $\xi$ as $\xi \rightarrow 0$, the exponent of the leading term needs to match a particular SLE exponent, as explained in each example.
    \item They must be real, non-negative, continuous, and differentiable, that is to say,
    \begin{equation}
        \rho_{m,m;n}(L, z) \geq 0 \quad\text{for} \quad z \in \mathbb{H} \quad \text{and} \quad L \in \mathbb{R}^+ \ ,
    \end{equation}
    with $\rho_{m,m;n}(L,z)$ continuous and differentiable in $z$ and $L$.
\end{itemize}
The second condition is a non-trivial requirement, since the solutions to \eqref{bpz2nd} and \eqref{bpz3rd} are in general complex valued.

\subsection{Insertion of 1-leg operators on the boundary} \label{sec:con-int}

In this section, we compute three-point functions in \eqref{11vv} that describe the densities \eqref{rho3pt} with $m=1$ and $n \in\{0, 2\}$. These densities are associated with an SLE curve from $x_1$ to $x_2$, where $x_1$ and $x_2$ are the insertion points of two 1-leg operators on the boundary (the real line). On a case-by-case basis, we will determine $G(\xi)$ in \eqref{11vv} as a linear combination of two fundamental solutions to \eqref{bpz2nd}. Since we will be comparing some of our computations with $\text{SLE}_\kappa$ computations, it is convenient to express the dimension $\Delta_{(2,1)}$ of $\phi_1(x)$ in terms of $\kappa$
\begin{equation}
    \Delta_{(2,1)} = \frac{3}{\kappa }-\frac{1}{2} \ .
\end{equation}

The operator $\phi_1(x)$ in \eqref{11vv} can be interpreted as the operator that changes the boundary condition from free to wired (or the other way around) \cite{Cardy92}. Therefore, in the examples below, we will assume that the interval $[x_1, x_2]$ is wired: every point on the real axis between $x_1$ and $x_2$ belongs to the same cluster. On the rest of the real line, $\mathbb{R} \setminus [x_1, x_2]$, we impose free boundary conditions.

\subsubsection{Cluster density}

We start with the density $\rho_{1,1; 0}(L, z)$, therefore we choose $\mathcal{O}(z)$ in \eqref{11vv} to be the spin field $\sigma(z)$ defined in \eqref{sig}. The dimension of $\sigma(z)$ is given by
\begin{equation} \label{dimsigk}
    \Delta_{(0, \frac{1}{2})} = \frac{1}{2}-\frac{1}{\kappa}-\frac{3 \kappa}{64} \ .
 \end{equation}

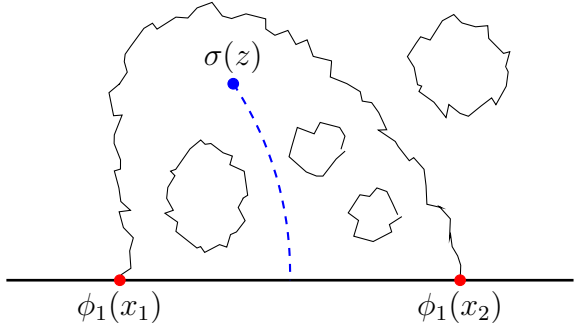
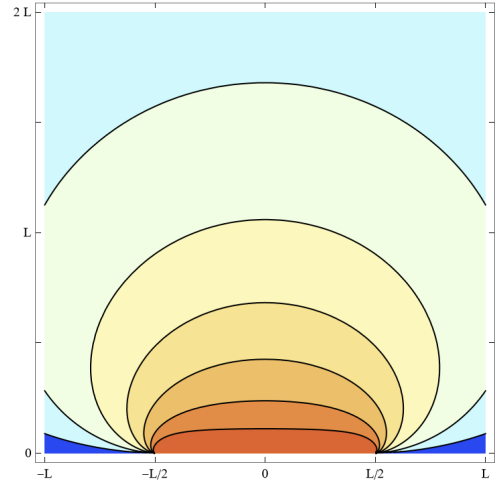
\begin{figure}[t]
\centering
\begin{subfigure}[b]{0.45\textwidth}
    \begin{tikzpicture}[x = 7.5 mm, y = 4.33 mm]
% 1L spin

\pgfmathsetseed{420}

\draw [line width=1pt] (-5,0) -- (5,0);
\draw[decorate, decoration={random steps,segment length=4pt, amplitude=3pt}] (-3,0) to[out angle=70, in angle=180-80, curve through = {(-3,6) (-2,8) (1,6)}] (3,0);

\draw[decorate, decoration={random steps,segment length=4pt, amplitude=3pt}] (-1.5,2.5) ellipse (5mm and 7mm);
\draw[decorate, decoration={random steps,segment length=4pt, amplitude=3pt}] (.5,4) circle (3mm);
\draw[decorate, decoration={random steps,segment length=4pt, amplitude=3pt}] (1.5,2) circle (3mm);
\draw[decorate, decoration={random steps,segment length=4pt, amplitude=3pt}] (3,6.5) circle (6mm);

\draw[dashed, thick, blue] (-1,6) to[curve through = {(0,.5)}] (0,0);

\filldraw[red] (-3,0) circle (2pt) node[black, anchor=north]{$\phi_1(x_1)$};
\filldraw[red] (3,0) circle (2pt) node[black, anchor=north]{$\phi_1(x_2)$};
\filldraw[blue] (-1,6) circle (2pt) node[black, anchor=south]{$\sigma(z)$};

\end{tikzpicture}
    \caption{The spin operator $\sigma$ in the bulk measures the density of FK-clusters that are connected to the boundary between the marked points.}
    \label{fig:1L_spin_a}
\end{subfigure} \hfill
\begin{subfigure}[b]{0.4\textwidth}
    \includegraphics[width=\textwidth]{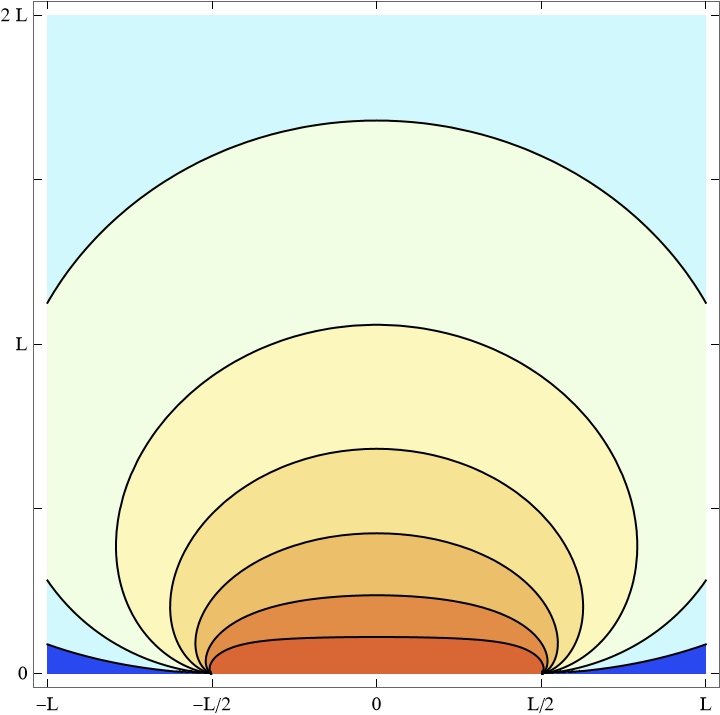}
    \caption{A contour plot of the corresponding density $\rho_{1,1;0}(L, z)$ with $\kappa=6$.}
    \label{fig:1L_spin_b}
\end{subfigure}
\caption{
A SLE curve starts and ends on the real line at $x_{1}$ and $x_2$, respectively. This corresponds to the insertion of the boundary 1-leg operators $\phi_1$.}
\label{fig:phiphisigma}
\end{figure}

We consider the interval $4<\kappa\leq8$, corresponding to non-simple $\text{SLE}_\kappa$ curves. Expressed as a cluster density, $\rho_{1,1; 0}(L, z)$ represents the (renormalized) probability that $z$ belongs to the (continuum) FK cluster containing the interval $[x_1,x_2]$, as shown in Figure \ref{fig:1L_spin_a}. 
In particular, this implies that the $\text{SLE}_\kappa$ curve corresponding to the boundary of the FK cluster containing $[x_1,x_2]$ passes to the left of $z$.
However, this alone does not guarantee that $z$ belongs to the cluster of $[x_1,x_2]$. This is the difference between $\rho_{1,1; 0}(L, z)$ and Schramm's left passage probability, which will be discussed in the next example.

With \eqref{dimsigk}, the two solutions to \eqref{bpz2nd} read
\begin{align}
g_1(\xi) =
 \left( 
1 + \frac{2-\xi}{2\sqrt{1- \xi} }
\right)^{\frac{\Delta_{(3,1)}}{2}}
\quad \text{and} \quad
g_2(\xi) =
 \left( 
1 - \frac{2-\xi}{2\sqrt{1- \xi} }
\right)^{\frac{\Delta_{(3,1)}}{2}}
\ ,
\label{gpm}
\end{align}
where
\begin{equation} \label{31k}
    \Delta_{(3,1)} = \frac{8}{\kappa }-1 \ .
\end{equation}
The density $\rho_{1,1;0}(L, z)$ is then a linear combination of the two solutions in \eqref{gpm}. To find the right combination, we consider the behavior of \eqref{gpm} as $\xi\rightarrow 0$:
\begin{subequations}
\begin{align}
    g_1(\xi)
    & \overset{\xi \rightarrow 0}\propto 1 + \ldots \\
    g_2(\xi)
    & \overset{\xi \rightarrow 0}\propto \xi^{\Delta_{(3,1)}} + \ldots \ .
\end{align}
\label{xi0}
\end{subequations}

We can think of the role of the two fundamental solutions as follows.
If $z \to x \in [x_1,x_2]$, the probability that $z$ belongs to the FK cluster containing $[x_1,x_2]$ should go to 1, corresponding to the asymptotic behavior of $g_1(\xi)$ as $\xi\to 0$.
If $z \to x \in \mathbb{R} \setminus [x_1,x_2]$, the same probability should go to zero with a precise speed of convergence, the same as that of the probability that an $\text{SLE}_\kappa$ curve from $x_1$ to $x_2$ comes close to $x$. The exponent determining the leading behavior of this probability is precisely $\Delta_{(3,1)}$, corresponding to the scaling dimension of the boundary 2-leg operator $\phi_2(x)$ and to the asymptotic behavior of $g_2$. Therefore, we conclude that
\begin{equation} \label{behavg}
G(\xi) \overset{\xi \rightarrow 0}\propto
\begin{cases}
    g_1(\xi) \quad \text{for} \quad &|x| \leq \frac{L}{2} \\
    g_2(\xi) \quad \text{for} \quad &|x| > \frac{L}{2} \ .
\end{cases}
\end{equation}
Combining the above observations with \eqref{11vv} and using \eqref{limrho} with $C=1$ and \eqref{cmmn}, we can write
\begin{equation} \label{110}
    \rho_{1,1;0}(L, z) = \frac{C_{1, 1;0}}{L^{2\Delta_{2,1}} (z-\bar{z})^{2 \Delta_{(0,\frac12)}}}
    \left( 1 + \frac{L^2 - 4z \bar z}{\sqrt{L^4  -4L^2(z^2 +\bar z^2) + 16z^2\bar z^2}} \right)^{\frac{\Delta_{(3,1)}}{2}} \ .
\end{equation}

For the percolation case, corresponding to $\kappa =6$, the density $\rho_{1,1;0}(L, z)$ reduces to the result of \cite{ksz06}. Note that the density diverges on the interval $[x_1,x_2]$ and vanishes on the real line outside that interval (see Figure \ref{fig:phiphisigma}).

The structure constant $C_{1, 1; 0}$ in \eqref{110} can be determined by considering the limit $z \rightarrow 0$ of $\rho_{1,1;0}(L, z)$. In this limit, we expect $\rho_{1,1;0}(L, z)$ to converge to the two-point function $\braket{ \phi_1(x_1) \phi_1(x_2) }$. Assuming that the fields are canonically normalized, that is, that their two-point functions have coefficient 1, we find $C_{1,1;0} = 1$.

\subsubsection{Schramm's left passage probability}

Schramm's left passage formula computes the probability that a chordal SLE curve starting and ending on the real line passes to the left of a point $z \in \mathbb{H}$. It is known that this probability formula can be obtained as a solution to the BPZ equation \eqref{bpz2nd} (for example, see Section 2.1 of \cite{Gamsa_2005}). Here, we review the discussion in detail. In CFT language, this probability corresponds to the ratio
\begin{equation} \label{slp}
    \left. \frac{\Braket{ \phi_1(x_1)\phi_1(x_2)\mathcal{O}(z) }_\mathbb{H}}{\Braket{ \phi_1(x_1)\phi_1(x_2) }_\mathbb{H}} \right\vert_{\Delta_\mathcal{O} = 0} \ ,
\end{equation}
where $\mathcal{O}(z)$ is a primary operator of dimension $0$. However, we stress here that $\mathcal{O}(z)$ is not the identity operator since, to arrive at Schramm's left passage, we need to require $\frac{\partial}{\partial z} \mathcal{O}(z) \neq 0$. For the case $\kappa = 8/3$, this operator $\mathcal{O}(z)$ is known as \emph{twist operator} \cite{gamsa2006correlation}.

We consider the BPZ equation \eqref{bpz2nd} with $\Delta_\mathcal{O} = 0$. In this case, it is convenient to choose $x_1=0$ and $x_2=\infty$. With these choices, two fundamental solutions to \eqref{bpz2nd} are given by
\begin{align}
\begin{split} \label{fslp}
    g_1(\xi) &= 1 \\
    g_2(\xi) &= u(\xi) \; {}_2F_1\left(\frac12,\frac4\kappa,\frac32; -u(\xi)^2\right) \ ,
\end{split}
\end{align}
with $u(\xi) = i\frac{\xi - 2}{\xi}$.

Since the SLE curve goes from $0$ to $\infty$, we have the physical boundary conditions
\begin{align} \label{gg}
\lim_{\stackrel{\xi \to 0}{z \to x \in \mathbb{R}}} G(\xi ) = 
\begin{cases}
    0 \quad\text{if} \quad x \leq 0 \\
    1 \quad\text{if} \quad x > 0
\end{cases}
\end{align}
Now observe that
\begin{equation}
    \lim_{\stackrel{\xi\to 0}{{z \to x \in \mathbb{R}}}}g_2(\xi ) = \operatorname{sgn}(x) \frac{\sqrt{\pi } \Gamma \left(\frac4\kappa-\frac{1}{2}\right)}{2 \Gamma \left(\frac4\kappa\right)} \ .
\end{equation}
This, together with \eqref{fslp}, leads to Schramm's left passage probability formula:
\begin{equation}
    G(\xi) = \frac12 - \frac{\Gamma\left(\frac4\kappa\right)}{\sqrt{\pi}\Gamma\left(\frac4\kappa-\frac{1}{2}\right)} u(\xi) \;
    {}_2F_1\left(\frac12,\frac4\kappa,\frac32; -u(\xi)^2\right) \ .
\end{equation}

\subsubsection{SLE path density}

\begin{figure}[t]
    \centering
    \begin{subfigure}[b]{0.45\textwidth}
        \begin{tikzpicture}[x = 7.5 mm, y = 4.33 mm]
% 1L 2L
\draw[line width=1pt] (-5,0) -- (5,0);
\draw[decorate, decoration={random steps,segment length=4pt, amplitude=3pt}] (-3,0) to[out angle=70, in angle=180-80, curve through = {(-3,6) (-2,8) (1,6)}] (3,0);

\filldraw[red] (-3,0) circle (2pt) node[black, anchor=north]{$\phi_1(x_1)$};
\filldraw[red] (3,0) circle (2pt) node[black, anchor=north]{$\phi_1(x_2)$};
\filldraw[blue] (-2,8) circle (2pt) node[black, anchor=south]{$\psi_2(z)$};

\end{tikzpicture}
        \caption{The 2-leg operator $\psi_2(z)$ in the bulk measures the density at $z$ of an SLE path starting and ending at the marked points on the boundary.}
        \label{fig:2L-1L}
    \end{subfigure} \hfill
    \begin{subfigure}[b]{0.4\textwidth}
        \includegraphics[width=\textwidth]{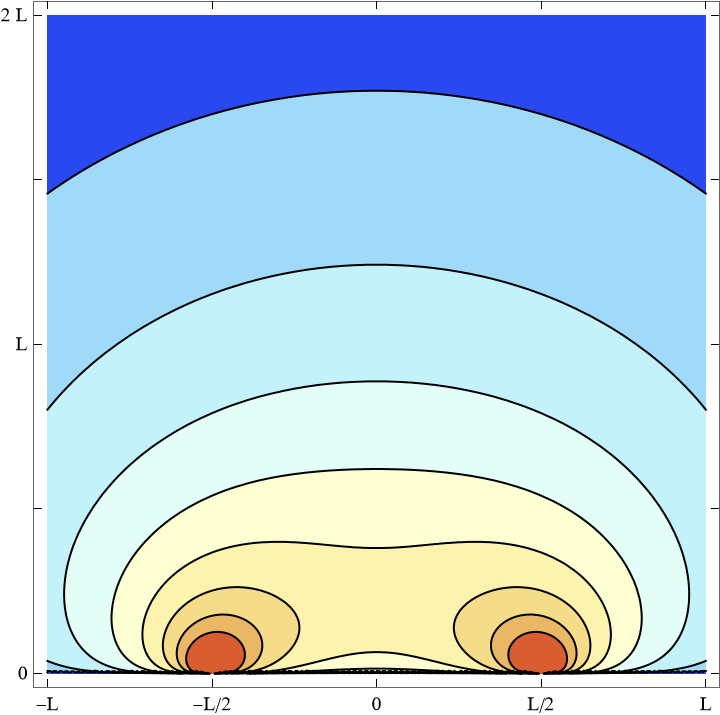}
        \caption{A contour plot of the corresponding density $\rho_{1,1;2}(L, z)$ with $\kappa = 6$.}
        \label{fig:2L_bdy}
    \end{subfigure}
    \caption{A SLE curve starts and ends on the real line at $x_{1}$ and $x_2$, respectively, while passing through the point $z$.}
    \label{2L_bdy}
\end{figure}
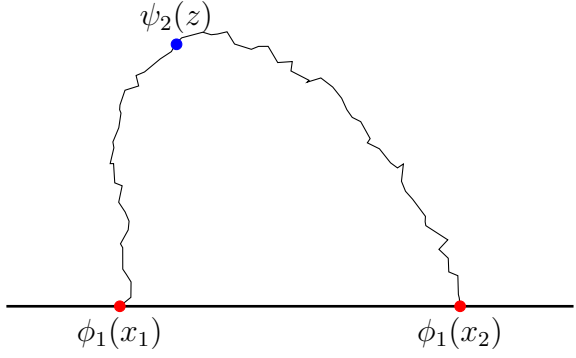
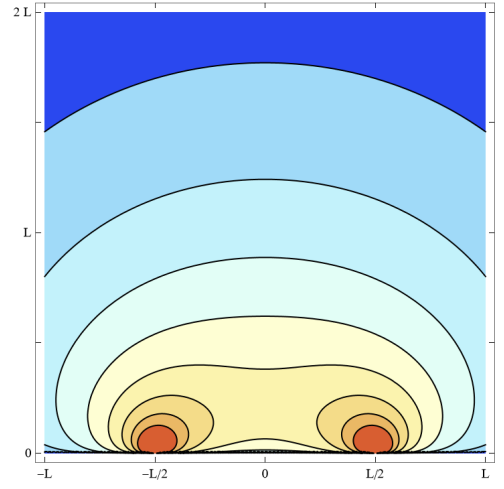

In this section, we consider the density $\rho_{1,1;2}(L, z)$ corresponding to the (renormalized) probability that an SLE curve in $\mathbb{H}$ from $x_1$ to $x_2$ goes through the bulk point $z \in \mathbb{H}$, as shown in Figure \ref{2L_bdy}. In CFT language, $\rho_{1,1;2}(L, z)$ can be expressed as the three-point function \eqref{11vv} where $\mathcal{O}(z)$ is identified with the bulk 2-leg operator $\psi_2(z)$ with dimension $\Delta_{(1,0)}$.
From Figure \ref{fig:2L-1L}, we see that the limit $z\rightarrow x \in \mathbb{R}$, with $x \neq x_1,x_2$, leads to the insertion of a boundary 2-leg operator at $x$. The dimension $\Delta_{(3,1)}$ of the corresponding operator, $\phi_2(x)$, is given by \eqref{31k}. Therefore, the physical solution should have the following behavior
\begin{align}
G(\xi) \overset{\xi \rightarrow 0}{\propto}
\xi^{\Delta_{(3,1)}} + \ldots \, .
\end{align}
Setting $\Delta_\mathcal{O}$ in \eqref{bpz2nd} to be $\Delta_{(1,0)}= \frac{8-\kappa }{16}$, we find that the desired solution is
\begin{align}
G(\xi) = \left(
\frac{\xi^2}{ 1-\xi}
\right)^{\frac{\Delta_{(3,1)}}{2}} \ .
\label{Gchordal}
\end{align}
Expressing $\xi$ as in \eqref{12xi}, then using \eqref{11vv} and \eqref{limrho}, we get
\begin{align} \label{112}
    \rho_{1,1; 2}(L, z) = \frac{C_{1,1 ;2}}
    {L^{2\Delta_{2,1}}(z-\bar{z})^{2\Delta_{(1,0)}}}
    \left(\frac{64 (z - \bar z)^2 L^2}{16 z^2\bar z^2+4 (z^2  + \bar z^2)L^2+L^4}\right)^{\frac{\Delta_{(3,1)}}{2}} \ .
\end{align}

The structure constant $C_{1, 1 ;2}$ will be determined in \eqref{listC} of Section \ref{sec:norm} by requiring that $\rho_{1,1; 2}(L, z)$ converges to the three-point function $\langle \phi_1(x_1)\phi_1(x_1)\phi_2(0) \rangle_{\mathbb{H}}$ as $z \rightarrow 0$, which corresponds to the (renormalized) probability that an SLE curve in $\mathbb{H}$ from $x_1$ to $x_2$ touches the origin.

Next, we will use \eqref{112} to obtain the SLE Green's function. As we will see in the next section, we will recover the expression derived rigorously in \cite{lgr15}, corroborating the validity of \eqref{112} and its derivation.

\subsubsection{SLE Green's function}

The expression for $\rho_{1,1; 2}(L, z)$ obtained above can be used to re-derive the 1-point chordal SLE$_\kappa$ Green's function, first derived in \cite{lgr15}, which is defined as
\begin{align}
    \mathbb{G}(z) = \lim_{\varepsilon \to 0} \varepsilon^{\frac{\kappa}{8} - 1} \mathbb{P}[ \operatorname{dist}(z,\gamma) \le  \varepsilon ] \, .
\end{align}
In CFT language, we can write the Green's function as follows:
\begin{align}
        \mathbb{G}(z) = \frac{ \Braket{ \psi_2(z) \phi_1(0) \phi_1(\infty) }_\mathbb{H} } {\Braket{\phi_1(0) \phi_1(\infty) }_{\mathbb{H}} } \, .
\end{align}
To compute the right-hand side of the above equation, we can use \eqref{Gchordal} and \eqref{11vv} and let $x_1 = 0$ and $x_2 \rightarrow \infty$, since we require the SLE$_\kappa$ curve to go from the origin to infinity. In this case, the cross-ratio becomes $\xi = \frac{z - \bar{z} }{z}$ and we find
\begin{align}
        \mathbb{G}(z)\propto (\Im{z})^{ \frac{(\kappa -8)^2}{8 \kappa } } \lvert z \rvert^{ 1 -\frac{8}{\kappa} } \, ,
\end{align}
as required \cite{lgr15}. As already remarked, this agreement with the mathematical literature supports the identification of the density in \eqref{112} with a solution to a BPZ equation obeying certain consistency conditions.

\subsection{Insertion of 2-leg operators on the boundary \label{sec:con-bd}}

In this section, we discuss the densities from~\eqref{rho3pt} with $m=2$, which correspond to the insertion of two boundary 2-leg operators $\phi_2$, with dimension given by~\eqref{31k}, and can be expressed in terms of the three-point function \eqref{22vv} with appropriate choices of $\mathcal{O}$, as explained below.

The densities $\rho_{2,2;n}(L, z)$ concern clusters or, more generally, SLE bubbles (loops) that touch two different points on the real axis, corresponding to the insertion of the two boundary 2-leg operators that appear in the three-point function \eqref{22vv}.
Such three-point functions satisfy the third-order BPZ equation \eqref{bpz3rd}.
The choice of $\mathcal{O}$ depends on the type of density under consideration (i.e., the observable of interest).
In the examples below, we will assume free boundary conditions. As mentioned earlier, in the Potts model, this is equivalent to having infinite temperature on the boundary.

\subsubsection{Cluster density}

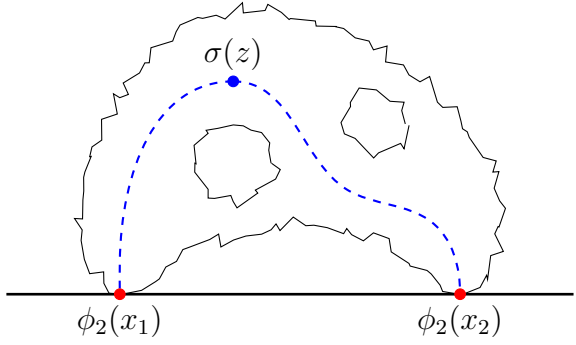
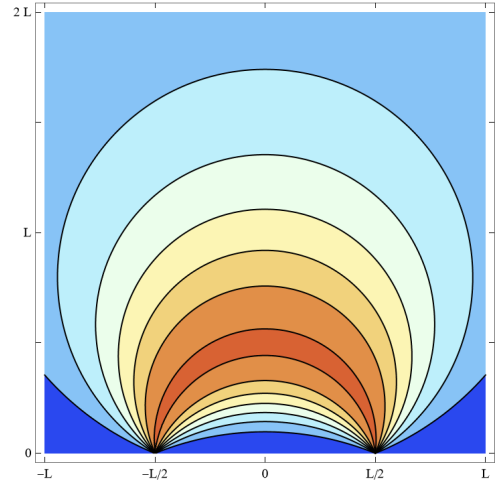
\begin{figure}[t]
    \centering
    \begin{subfigure}[b]{0.45\textwidth}
        \begin{tikzpicture}[x = 7.5 mm, y = 4.33 mm]
% 2L spin

\pgfmathsetseed{25939}
\draw[line width=1pt] (-5,0) -- (5,0);

\draw[decorate, decoration={random steps,segment length=4pt, amplitude=3pt}] (-3,0) to[out angle=180-30, in angle=30, curve through = {(-3,6) (-2,8) (2,7)}] (3,0);
\draw[decorate, decoration={random steps,segment length=4pt, amplitude=3pt}] (-3,0) to[out angle=30, in angle=180-30, curve through = {(-2,1) (1,2)}] (3,0);

\draw[decorate, decoration={random steps,segment length=4pt, amplitude=3pt}] (1.5,5.2) circle (4mm);
\draw[decorate, decoration={random steps,segment length=4pt, amplitude=3pt}] (-1,4) circle (5mm);

\draw[dashed, thick, blue] (-1,6.5) to[out angle=180, curve through = {(-3,1)}] (-3,0);
\draw[dashed, thick, blue] (-1,6.5) to[out angle=0, in angle=90, curve through = {(1,3.2) (2.5,2.2)}] (3,0);

\filldraw[red] (-3,0) circle (2pt) node[black, anchor=north]{$\phi_2(x_1)$};
\filldraw[red] (3,0) circle (2pt) node[black, anchor=north]{$\phi_2(x_2)$};
\filldraw[blue] (-1,6.5) circle (2pt) node[black, anchor=south]{$\sigma(z)$};

\end{tikzpicture}
        \caption{The spin operator $\sigma$ in the bulk measures the density of an FK-cluster connected to the marked points $x_1$ and $x_2$ on the boundary.}
        \label{fig:2L-spin}
    \end{subfigure} \hfill
    \begin{subfigure}[b]{0.4\textwidth}
        \includegraphics[width=\textwidth]{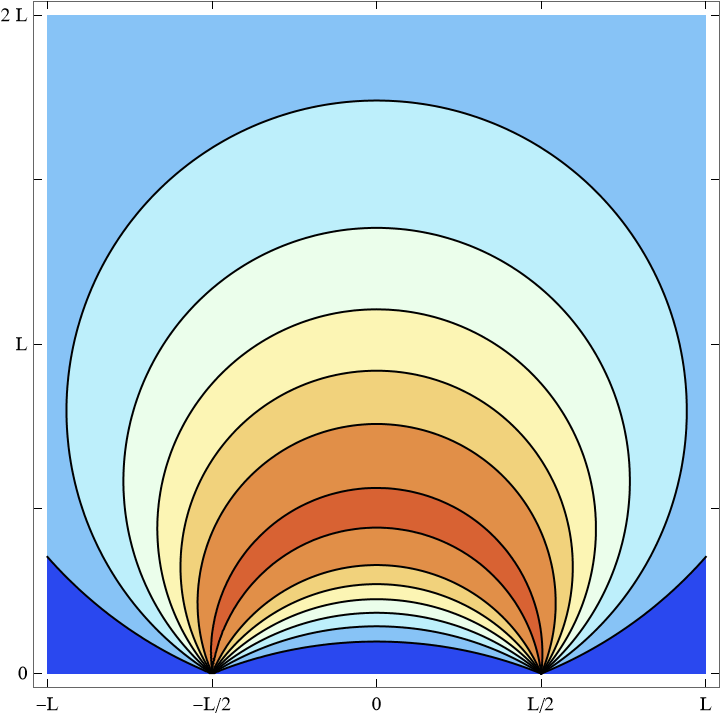}
        \caption{A corresponding contour plot of the density $\rho_{2,2; 0}(L, z)$ with $\kappa = 8/3$.}
        \label{fig:2L_con}
    \end{subfigure}
    \caption{SLE path touching the boundary, corresponding to the insertion of two boundary 2-leg operators and one spin field in the bulk.}
    \label{fig:2L}
\end{figure}

If we choose $\mathcal{O}(z)$ in~\eqref{22vv} to be the spin field $\sigma(z)$ defined in~\eqref{sig}, the density $\rho_{2,2;0}(L, z)$ corresponds to the event that $z$ is contained inside an SLE bubble that touches the real line twice, but not inside a smaller loop (see Figure~\ref{fig:2L-spin}). The lattice interpretation is the probability that a point $z$ belongs to an FK cluster that touches the real line at two distinct locations.

Now observe that letting $z \rightarrow x \in\mathbb{R}$ corresponds to inserting a boundary 2-leg operator at $x$, $\phi_2(x)$, with dimension $\Delta_{(3,1)}$. Therefore, we expect the function $H$ in~\eqref{22vv} to behave as
\begin{equation} \label{be110}
    H(\xi) \overset{\xi \rightarrow 0} \propto \xi^{\Delta_{(3,1)}} + \ldots \ .
\end{equation}
A solution to \eqref{bpz3rd} that behaves as in~\eqref{be110} has a very simple expression, given by
\begin{equation}
    H(\xi) = \left( \frac{\xi^2}{1-\xi} \right)^{\frac{\Delta_{(3,1)}}{2}} \ .
\end{equation}
Note that this is the same function as in~\eqref{Gchordal} and, indeed, sending $z \to x \in \mathbb{R}$ in Figures~\ref{fig:2L-1L} and~\ref{fig:2L-spin} has the same effect: it corresponds to the insertions of 2 legs at $x$.

With \eqref{rho3pt}, \eqref{22vv} and \eqref{limrho}, we can then write
\begin{equation} \label{220}
    \rho_{2,2;0}(L, z) =
    \frac{C_{2,2;2}}{L^{2\Delta_{3,1}}{(z-\bar{z})}^{2\Delta_{(0,\frac12)}}}
    \left(\frac{64 (z- \bar z)^2 L^2}{16z^2\bar z^2 +4 (z^2 + \bar z^2)L^2+L^4}\right)^{\frac{\Delta_{(3,1)}}{2}} \ ,
\end{equation}
where $C_{2,2;2}$ is given in~\eqref{listC} below.
The presence of the structure constant $C_{2,2;2}$ in~\eqref{220} follows from the requirement that, as $z \rightarrow 0$, $\rho_{2,2;0}(L, z)$ should converge to the three-point function $\braket{ \phi_2(x_1)\phi_2(x_2)\phi_2(0) }_{\mathbb{H}}$. We will explain how to determine this three-point function in Section \ref{sec:norm} below.

Note that the density $\rho_{2,2; 0}(L, z)$ decreases as $|z| \rightarrow \infty$ and increases near the points where the boundary operators $\phi_2$ are inserted, as expected (see Figure \ref{fig:2L_con}).

\subsubsection{Density of the outer boundary}

\begin{figure}[t]
\centering
    \begin{subfigure}[b]{0.45\textwidth}
    \begin{tikzpicture}[x = 7.5 mm, y = 4.33 mm]
% 2L 2L

\pgfmathsetseed{25939}
\draw [line width=1pt] (-5,0) -- (5,0);

\draw[decorate, decoration={random steps,segment length=4pt, amplitude=3pt}] (-3,0) to[out angle=180-30, in angle=30, curve through = {(-3,6) (-2,8) (2,7)}] (3,0);
\draw[decorate, decoration={random steps,segment length=4pt, amplitude=3pt}] (-3,0) to[out angle=30, in angle=180-30, curve through = {(-2,1) (0,2.5) (1,2)}] (3,0);

\filldraw[red] (-3,0) circle (2pt) node[black, anchor=north]{$\phi_2(x_1)$};
\filldraw[red] (3,0) circle (2pt) node[black, anchor=north]{$\phi_2(x_2)$};
\filldraw[blue] (0,2.5) circle (2pt) node[black, anchor=south]{$\psi_2(z)$};
%2.1
\end{tikzpicture}
    \caption{The density of the bubble boundary is calculated by inserting one bulk 2-leg operator, here shown on the lower portion of the boundary.}
    \label{fig:2L-2L-in}
    \end{subfigure} \hfill
    \begin{subfigure}[b]{0.4\textwidth}
    \includegraphics[width=\textwidth]{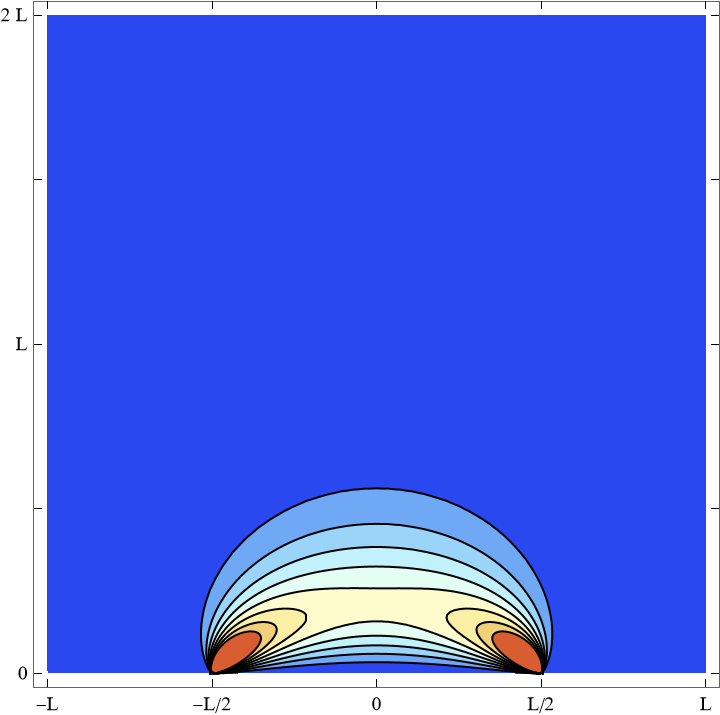}
    \caption{A corresponding contour plot of the density $\rho^\text{lower}_{2,2; 2}(L, z)$ with $\kappa = 8/3$.}
    \label{fig:2L-2L-den}
    \end{subfigure}
    \caption{The SLE bubble is touching the boundary in two places, corresponding to the insertion of two boundary 2-leg operators.}
    \label{fig:2l_2l-2}
\end{figure}
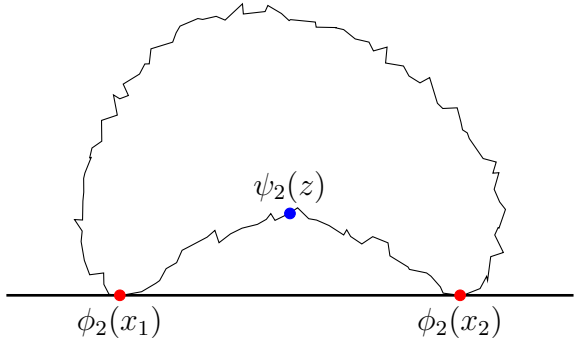
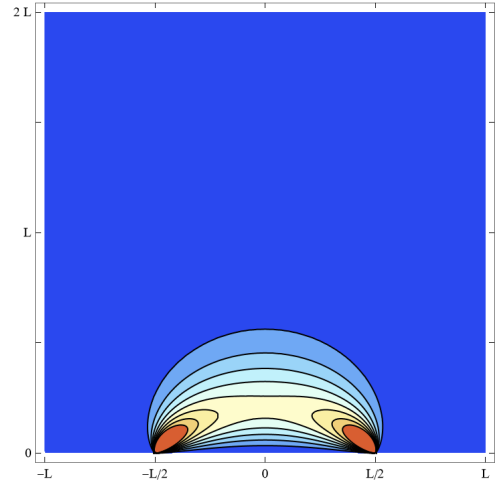

We proceed with a calculation that has not previously appeared in the literature, namely that of the density $\rho_{2,2;2}(L, z)$ of an SLE bubble anchored to two points on the boundary (see Figure~\ref{fig:2L-2L-in}), which can be written as
\begin{align} \label{rho222}
    \rho_{2,2; 2}(L, z) = \frac{C_{2,2; 2}}{L^{2\Delta_{3,1}}(z-\bar{z})^{2\Delta_{(1,0)}}} H(\xi) \ .
\end{align}

To compute the function $H(\xi)$, we consider the three-point function \eqref{22vv} with $\mathcal{O}(z)=\psi_2(z)$.
This three-point function satisfies a third order BPZ equation of type \eqref{bpz3rd}.
The two relevant solutions are
\begin{subequations}
\begin{align}
h_1(\xi) &=
\left(\frac{\xi}{\xi - 1}\right)^{\frac{8}{\kappa }-1}  \, _2F_1\left(1-\frac{8}{\kappa },1-\frac{8}{\kappa },2-\frac{16}{\kappa };\xi \right)
\label{hnor} \\
h_2(\xi) &=
\xi^{\frac{24}{k} -2} (\xi - 1)^{1-\frac{8}{\kappa }} \, _2F_1\left(\frac8\kappa, \frac8\kappa, \frac{16}{\kappa};\xi \right)
\ .
\label{hnor2}
\end{align}
\end{subequations}
In the limit $\xi \rightarrow 0$, we see that $h_1(\xi) \propto \xi^{\Delta_{(3,1)}}$ and $h_2(\xi) \propto \xi^{\Delta_{(5,1)}}$, which correspond to the asymptotic behaviors that we expect from the solutions of~\eqref{bpz3rd}.

Appropriate combinations of $h_1$ and $h_2$ describe the density of the boundary of the bubble, where upper and lower portions must be described separately.
For example, the combination corresponding to the lower portion of the boundary of the bubble should be a combination of the following solutions:
\begin{align} \label{2bc}
    H(\xi)\overset{\xi\rightarrow 0} \propto
    \begin{cases}
        h_1(\xi) = \xi^{\Delta_{(3,1)}}+\ldots \quad\text{if}\quad z \to x\in \left(x_1,x_2\right) \\
        h_2(\xi) = \xi^{\Delta_{(5,1)}}+\ldots \quad\text{if}\quad z \to x \in \mathbb{R}\setminus\left[x_1,x_2\right] \ .
    \end{cases}
\end{align}
This can be understood from Figure \ref{fig:2L-2L-in}. If $\psi_2$ describes the density of the lower branch and approaches the real line between the insertion points $x_1$ and $x_2$, the bulk 2-leg operator then creates two excursions starting from the limit point on the real line. This is exactly the behavior of $\phi_2$ with dimension $\Delta_{(3,1)}$. If $\psi_2$ approaches the real line outside the interval of length $L$, $\psi_2$ forces the upper portion of the bubble to also touch the real line. This effectively creates 4 excursions from the limit point, corresponding to $\phi_4$ with dimension $\Delta_{(5,1)}$. If we were to instead measure the density of the upper portion, the scaling behavior of our choice for $H$ would be flipped
\begin{align} \label{22bc}
    H(\xi)\overset{\xi\rightarrow 0} \propto
    \begin{cases}
        h_1(\xi) = \xi^{\Delta_{(3,1)}}+\ldots \quad\text{if}\quad z \to x \in \mathbb{R}\setminus\left[x_1,x_2\right] \\
        h_2(\xi) = \xi^{\Delta_{(5,1)}}+\ldots \quad\text{if}\quad z \to x\in \left(x_1,x_2\right) \ .
    \end{cases}
\end{align}
For example, using \eqref{2bc}, the density of the lower portion is given by
\begin{align} \label{lowerportion}
    \rho^\text{lower}_{2,2; 2}(L, z) \propto \frac{1}{L^{2\Delta_{3,1}}(z-\bar{z})^{2\Delta_{(1,0)}}} 
    \begin{cases}
        c_1 h_1(\xi) + c_2 h_2(\xi) & |z| < L/2 \\
        h_2(\xi) & |z| \ge L/2 \ .
    \end{cases}
\end{align}
Assuming that the density $\rho^\text{lower}_{2,2; 2}(L, z)$ corresponds to a well-defined correlation function within a consistent conformal field theory, the constants $c_1$ and $c_2$ are uniquely determined as solutions to the crossing-symmetry equation. Specifically, we utilize a constraint analogous to equation (3.32) of \cite{lew92}, which is derived by requiring that the two distinct regimes outlined in \eqref{lowerportion} map consistently onto one another via analytic continuation in the complex parameter $\xi$.
This implies algebraic conditions, which can be solved to obtain the constants
\begin{align}
\begin{split}
    c_1 &= e^{-16 \pi i / \kappa} \left( \frac{2\pi \kappa}{\kappa-16} \right) \frac{\Gamma(16/\kappa)^2}{\Gamma(8/\kappa)^4} \\
    c_2 &= \frac{e^{-24 \pi i / \kappa}}{\cos(8\pi/\kappa)}.
\end{split}
\end{align}
The contour plot of one such example is shown in Figure \ref{fig:2L-2L-den}.

In the scaling limit, the boundary of the hull of a critical percolation cluster (see Figure \ref{fjords}) is also described by an SLE curve with $\kappa=8/3$. Therefore, for $\kappa=8/3$, we can think of $\rho_{2,2;2}(L, z)$ as the density at $z$ of the boundary of the hull of a percolation cluster anchored to $x_1=-L/2$ and $x_2=L/2$ on the real line.

\begin{figure}[t]
    \centering
    \includegraphics[scale = 0.7]{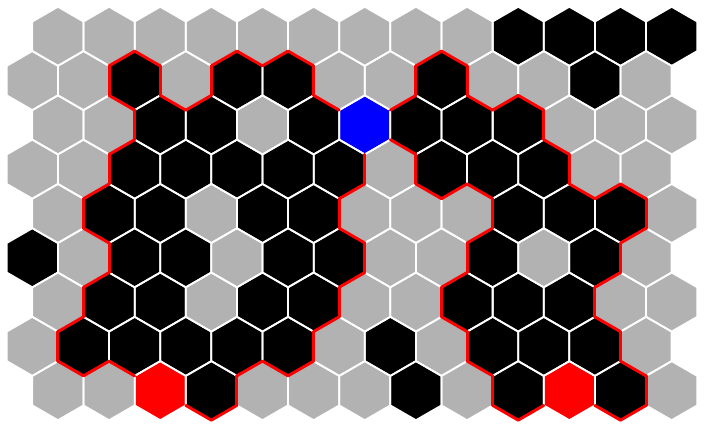}
    \caption{A site percolation lattice configuration with two clusters pinned to the real line (at the red hexagons) and a pivotal hexagon (in blue). In the scaling limit, the red outer boundaries of the clusters converge to two SLE bubbles anchored to the real line, as shown in Figure \ref{fig:4L-4L}.}
    \label{fig:4l_lattice}
\end{figure}

\subsubsection{Density of pivotal points}

\begin{figure}[t]
    \centering
    \begin{subfigure}[b]{0.45\textwidth}
        \begin{tikzpicture}[x = 7.5 mm, y = 4.33 mm]
% 2L 4L
\draw[line width=1pt] (-5,0) -- (5,0);

\draw[decorate, decoration={random steps,segment length=4pt, amplitude=3pt}] (-3,0) to[out angle=140, in angle=110, curve through = {(-3,6) }] (0,4);
\draw[decorate, decoration={random steps,segment length=4pt, amplitude=3pt}] (-3,0) to[out angle=40, in angle=-130, curve through = {(-1,2)}] (0,4);

\draw[decorate, decoration={random steps,segment length=4pt, amplitude=3pt}] (3,0) to[out angle=50, in angle=80, curve through = {(3,6)}] (0,4);
\draw[decorate, decoration={random steps,segment length=4pt, amplitude=3pt}] (3,0) to[out angle=140, in angle=-60, curve through = {(1,2)}] (0,4);

%[decorate, decoration={random steps,segment length=4pt, amplitude=3pt}]

\filldraw[red] (-3,0) circle (2pt) node[black, anchor=north]{$\phi_2(x_1)$};
\filldraw[red] (3,0) circle (2pt) node[black, anchor=north]{$\phi_2(x_2)$};
\filldraw[blue] (0,4) circle (2pt) node[black, anchor=east]{$\psi_4(z)$};

\end{tikzpicture}
        \caption{The 4-leg operator $\psi_4(z)$ inserts four legs at the bulk point $z$, produced by two bubbles meeting at that point.}
        \label{fig:4L_4L_SLE}
    \end{subfigure} \hfill
    \begin{subfigure}[b]{0.4\textwidth}
        \includegraphics[width=\textwidth]{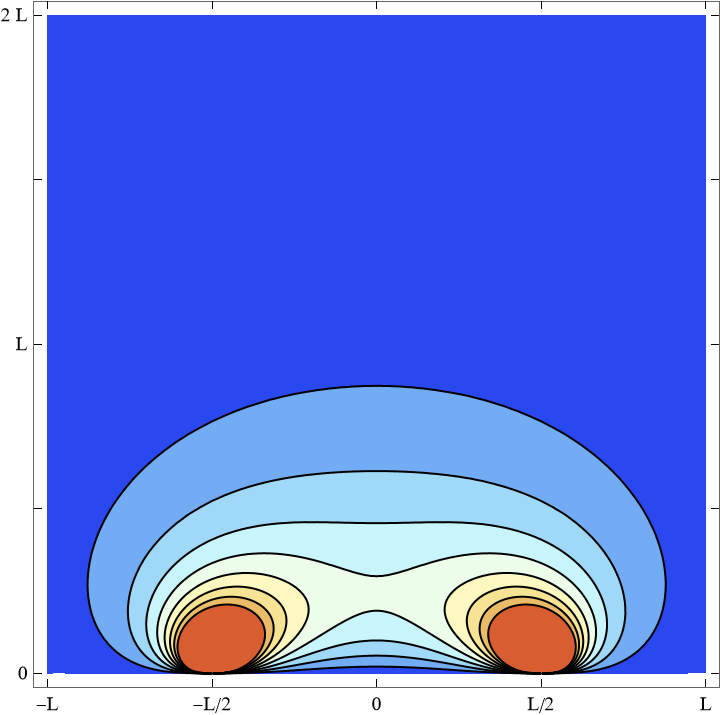}
        \caption{A corresponding contour plot of the density $\rho_{2,2;4}(L, z)$ with $\kappa = 6$.}
        \label{fig:4L_con}
    \end{subfigure}
    \caption{Two paths touch the real line, corresponding to inserting the boundary 2-leg operators $\phi_2$.}
    \label{fig:4L-4L}
\end{figure}
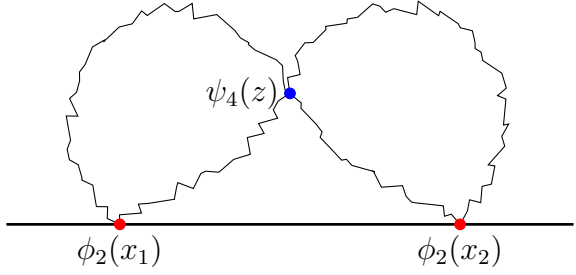
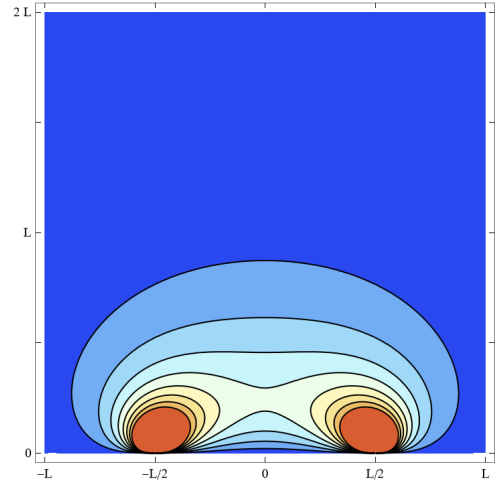

We now present another novel result.
We consider two separate clusters (and the corresponding SLE bubbles in the continuum limit) pinned to two distinct points on the real line.
If the two clusters/bubbles ``touch'', they create a \emph{pivotal point} --- see Figure \ref{fig:4l_lattice} for an example in the case of percolation.
The density of these pivotal points can be expressed using \eqref{22vv} chosing $\mathcal{O}(z)=\psi_4(z)$, where $\psi_4$ denotes the 4-leg operator in the bulk with dimension
\begin{equation}
    \Delta_{(2,0)} = -\frac{\kappa }{16}+\frac{3}{\kappa }+\frac{1}{2} \ .
\end{equation}

As $\xi \rightarrow 0$, the function $H(\xi)$ in \eqref{22vv} should have the following behavior:
\begin{equation}
H(\xi)
\overset{\xi \rightarrow 0}{\propto}
\xi^{\Delta_{(5,1)}}
+\ldots
\ .
\end{equation}
We find that a solution to the differential equation \eqref{bpz3rd} with the correct behavior is
\begin{equation}
H(\xi)
= \left(
\frac{\xi^2}{1 - \xi}
\right)^{\Delta_{(5,1)}}
\ .
\label{Hpi}
\end{equation}
Combining \eqref{Hpi} and \eqref{22vv}, we find
\begin{align}
\rho_{2,2;4}(L, z)
=
\frac{C_{2, 2 ;4}}{
\Im(z)^{2\Delta_{(2,0)}}}
\left(\frac{64 (z - \bar z)^2 L^2}{16 z^2\bar z^2 + 4(z^2 + \bar z^2) L^2+L^4}\right)
^{\frac{\Delta_{(5,1)}}{2}}
\ ,
\label{224}
\end{align}
where $C_{2, 2 ;4}$ is given in \eqref{listC} below.

Figure \ref{fig:4L_con} shows that the density $\rho_{2,2;4}(L, z)$ is high near the insertion points of the boundary 2-leg operators $\phi_2$ and vanishes as $z \rightarrow x \in\mathbb{R}$ with $x \neq x_1,x_2$. This behavior is consistent with the fact that a pivotal point on the boundary corresponds to a three-arm event on the boundary. The boundary three-arm exponent, which determines how fast the probability of a three-arm event on the boundary goes to zero in the scaling limit, is universal (i.e., the same for FK models with different values of $Q$) and equals 2 (see~\cite{Copin2021}), which implies that, in the scaling limit, there is zero probability to see a three-arm event anywhere on the boundary.

The presence of the structure constant $C_{2, 2 ;4}$ follows from the requirement that, as $z \rightarrow 0$, $\rho_{2,2;4}(L, z)$ converges to the three-point function $\braket{ \phi_2(x_1)\phi_2(x_2)\phi_4(0) }$.
In Section~\ref{sec:norm}, we will determine this three-point function by analyzing the boundary four-point function of $\phi_2(x)$.

\subsection{Fixing the normalization {\label{sec:norm}}}

In this section, we discuss how to fix the normalization of the densities $\rho_{m, m; n}(L, z)$. As previously mentioned, this requires computing the structure constants $C_{m,m; j}$ in \eqref{cmmn}, and it turns out that there are only three of these structure constants that are non-trivial and are relevant to the the densities $\rho_{m, m; n}(L, z)$ computed in our examples. They are given by
\begin{subequations}
\begin{align}
C_{1, 1 ;2} &=
\sqrt{\frac{\pi  \left(\csc \left(\frac{4 \pi }{\kappa }\right)+\csc \left(\frac{12 \pi }{\kappa
   }\right)\right) \Gamma \left(1-\frac{8}{\kappa }\right) \Gamma \left(2-\frac{8}{\kappa
   }\right)}{\Gamma \left(2-\frac{12}{\kappa }\right) \Gamma \left(1-\frac{4}{\kappa }\right)^2
   \Gamma \left(\frac{4}{\kappa }\right)}}
\label{c112}
\\
C_{2, 2 ;2} &=
\sqrt{-\frac{\pi  \left(\csc \left(\frac{8 \pi }{\kappa }\right)+\csc \left(\frac{16 \pi
   }{\kappa }\right)\right) \sec \left(\frac{8 \pi }{\kappa }\right) \Gamma
   \left(2-\frac{8}{\kappa }\right) \Gamma \left(\frac{4}{\kappa }\right) \Gamma
   \left(\frac{\kappa -8}{\kappa }\right) \Gamma \left(\frac{2 (\kappa -6)}{\kappa
   }\right)}{2 \Gamma \left(\frac{8}{\kappa }\right)^2 \Gamma \left(\frac{2 (\kappa
   -8)}{\kappa }\right)^2 \Gamma \left(\frac{\kappa -4}{\kappa }\right)^2}}
\label{c222}
\\
C_{2, 2 ;4} &=
\sqrt{\frac{(\kappa -8) \left(2 \cos \left(\frac{8 \pi }{\kappa }\right)+1\right)^2 \Gamma
   \left(\frac{\kappa -12}{\kappa }\right) \Gamma \left(\frac{2 (\kappa -6)}{\kappa
   }\right)}{(\kappa -16) \left(2 \cos \left(\frac{8 \pi }{\kappa }\right)+2 \cos
   \left(\frac{16 \pi }{\kappa }\right)+1\right) \Gamma \left(2-\frac{20}{\kappa
   }\right) \Gamma \left(\frac{\kappa -4}{\kappa }\right)}}
\ .
\end{align}
\label{listC}
\end{subequations}
For percolation, corresponding to $\kappa = 6$, we find
\begin{subequations}
\begin{align}
C_{1, 1 ;2}^{\kappa =6} &=
\frac{\sqrt{-\Gamma \left(-\frac{1}{3}\right)}}{\Gamma \left(\frac{1}{3}\right)}= 0.752361\ldots
\label{c1126}
\\
C_{2, 2 ;2}^{\kappa =6} &=
\sqrt{-\frac{96 \pi ^{7/2}}{\Gamma \left(-\frac{2}{3}\right) \Gamma \left(\frac{1}{6}\right)^3 \Gamma
   \left(\frac{1}{3}\right)^2}}
   =1.02993\ldots
\\
C_{2, 2 ;4}^{\kappa =6}&=
2^{\frac34}3^{-\frac58} \sqrt[4]{\frac{\pi }{5}}
=0.56785\ldots
\ .
\end{align}
\end{subequations}
In particular, $C_{2, 2 ;2}^{\kappa =6}$ perfectly matches with the same result for percolation in \cite{skz07}. Moreover, the numerical value of $C_{2, 2 ;2}^{\kappa =6}$ is in excellent agreement with the Monte Carlo result of \cite{ksz06}, which is $C_{2, 2;2}^{\kappa =6} = 1.030 \pm 0.001$.

The structure constants in \eqref{listC} can be extracted from the two boundary four-point functions $\braket{ \phi_1(x_1)\phi_1(x_2)\phi_1(x_3)\phi_1(x_4) }$ and $\braket{ \phi_2(x_1)\phi_2(x_2)\phi_2(x_3)\phi_2(x_4) }$, which can be computed as linear combinations of solutions to the BPZ equations. We now discuss their computation.

\subsubsection{Four-point function of boundary 1-leg operators}

\begin{figure}
\centering
    \begin{tikzpicture}[x = 7.5 mm, y = 4.33 mm]
% 2L 2L

\pgfmathsetseed{25939}
\draw [line width=1pt] (-15,0) -- (5,0);

\draw[decorate, decoration={random steps,segment length=4pt, amplitude=3pt}] (-3,0) to[out angle=180-30, in angle=30, curve through = {(-3,3) (-2,4) (2,3)}] (3,0);
%\draw[decorate, decoration={random steps,segment length=4pt, amplitude=3pt}] (-3,0) to[out angle=30, in angle=180-30, curve through = {(-2,1) (1,2)}] (3,0);

\draw[decorate, decoration={random steps,segment length=4pt, amplitude=3pt}] (-12,0) to[out angle=180-30, in angle=30, curve through = {(-12,3) (-11,4) (-7,3)}] (-6,0);
%\draw[decorate, decoration={random steps,segment length=4pt, amplitude=3pt}] (-12,0) to[out angle=30, in angle=180-30, curve through = {(-11,1) (-8,2)}] (-6,0);

%\draw[decorate, decoration={random steps,segment length=4pt, amplitude=3pt}] (-6,0) to[out angle=180-30, in angle=30, curve through = {(-6,6) (-2,9) (2,7)}] (6,0);

%\draw[decorate, decoration={random steps,segment length=4pt, amplitude=3pt}] (-6,0) to[out angle=180-30, in angle=30, curve through = {(-4,4) (-2,6) (2,4)}] (6,0);

\filldraw[red] (-3,0) circle (2pt) node[black, anchor=north]{$\phi_1(x_3)$};
\filldraw[red] (3,0) circle (2pt) node[black, anchor=north]{$\phi_1(x_4)$};

\filldraw[red] (-12,0) circle (2pt) node[black, anchor=north]{$\phi_1(x_1)$};
\filldraw[red] (-6,0) circle (2pt) node[black, anchor=north]{$\phi_1(x_2)$};

\end{tikzpicture}
    \caption{The two pairs of boundary 1-leg operators insert two non-intersecting SLE paths on the upper half-plane that start and end on the real axis. There are two possible configurations, the one shown in the figure and the one obtained by exchanging the positions of the points $x_2$ and $x_4$.}
    \label{fig:1L-1L-4}
\end{figure}
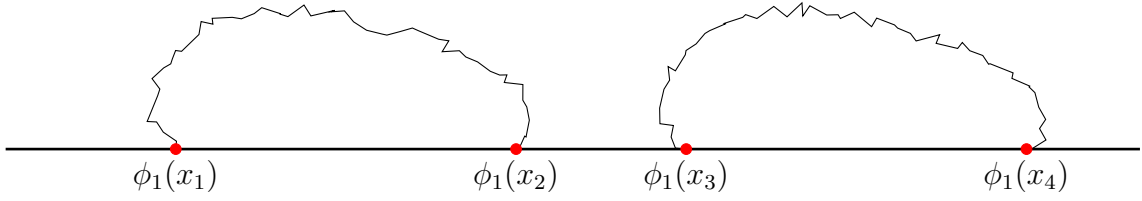

The four-point function of the boundary 1-leg operator, i.e. $\braket{ \phi_1(x_1)\phi_1(x_2)\phi_1(x_3)\phi_1(x_4) }$, corresponds to two SLE curves that connect the points $x_1$, $x_2$, $x_3$, and $x_4$, as shown in Figure~\ref{fig:1L-1L-4}. Global conformal invariance dictates that this four-point function takes the form
\begin{equation}
\Braket{ \phi_1(x_1)\phi_1(x_2)\phi_1(x_3)\phi_1(x_4) }
= \frac{G(\xi )}{
(x_1 - x_2)^{2\Delta_{(2,1)}}
(x_3 - x_4)^{2\Delta_{(2,1)}}
} \ ,
\end{equation}
where the cross-ratio $\xi$ is given in \eqref{xi} with $z$ and $\bar z$ replaced by $x_3$ and $x_4$, respectively.

The function $G(\xi )$ is known \cite{fms97} --- it is the solution to the second-order BPZ equation \eqref{bpz2nd} --- and we will refrain from writing it down explicitly. Now, the geometry of the SLE curves in Figure~\eqref{fig:1L-1L-4} implies that, in the limit $x_2,x_3 \rightarrow x$ or equivalently $\xi \rightarrow 1$, $G(\xi )$ must behave as
\begin{equation} \label{g31}
    G(\xi ) \overset{\xi \rightarrow 1}{ \propto} (1 - \xi)^{\Delta_{(3, 1)}} + \ldots \ .
\end{equation}
In other words, as $x_2,x_3 \rightarrow x$, the two SLE curves of Figure~\eqref{fig:1L-1L-4} should combine into one curve that goes from $x_1$ to $x_4$ and touches the real line at $x$. Therefore, the relevant solution of~\eqref{bpz2nd} is of the form~\eqref{2bpz}, with the two fundamental solutions of~\eqref{bpz2nd} chosen so that their linear combination obeys \eqref{g31}. Next, we recall that, in addition to \eqref{asymp}, for this boundary four-point function, we have
\begin{subequations}
\begin{align}
    g^{(1)}_1(\xi) &\overset{\xi \rightarrow 1}\propto 1 + \ldots \\
    g^{(1)}_2(\xi) &\overset{\xi \rightarrow 1}\propto (1 - \xi)^{\Delta_{(3,1)}} + \ldots \ .
\end{align}
\label{gsymp}
\end{subequations}
Therefore, using the transformation \eqref{F2b} with \eqref{g31} and \eqref{gsymp}, we find that the required combination is
\begin{align}
\begin{split} \label{com1l}
    G( \xi) &= g^{(0)}_1(\xi) - \frac{F_{11}}{F_{21}}g^{(0)}_2(\xi) \\
    &=  \left( F_{12} - \frac{F_{11}F_{22}}{F_{21}} \right)g^{(1)}_2(\xi)
\end{split}
\end{align}
with
\begin{equation} \label{cst1}
 F_{12} - \frac{F_{11}F_{22}}{F_{21}} =
\frac{\pi  \left(\csc \left(\frac{4 \pi }{\kappa }\right)+\csc \left(\frac{12 \pi }{\kappa
   }\right)\right) \Gamma \left(1-\frac{8}{\kappa }\right) \Gamma \left(2-\frac{8}{\kappa
   }\right)}{\Gamma \left(2-\frac{12}{\kappa }\right) \Gamma \left(1-\frac{4}{\kappa }\right)^2
   \Gamma \left(\frac{4}{\kappa }\right)} \ ,
\end{equation}
where the coefficient of $g^{(0)}_1(\xi)$ is the identity structure constant, which is canonically normalized to be 1.
From the OPE, the quantity \eqref{cst1} can be interpreted as the square of the structure constant $C_{1,1; 2}$, which determines the value of $C_{1,1; 2}$ given in \eqref{c112}.

\subsubsection{Four-point function of boundary 2-leg operators}

\begin{figure}
\centering
    \begin{tikzpicture}[x = 7.5 mm, y = 4.33 mm]
% 2L 2L

\pgfmathsetseed{25939}
\draw [line width=1pt] (-15,0) -- (5,0);

\draw[decorate, decoration={random steps,segment length=4pt, amplitude=3pt}] (-3,0) to[out angle=180-30, in angle=30, curve through = {(-3,3) (-2,4) (2,3)}] (3,0);
\draw[decorate, decoration={random steps,segment length=4pt, amplitude=3pt}] (-3,0) to[out angle=30, in angle=180-30, curve through = {(-2,1) (1,2)}] (3,0);

\draw[decorate, decoration={random steps,segment length=4pt, amplitude=3pt}] (-12,0) to[out angle=180-30, in angle=30, curve through = {(-12,3) (-11,4) (-7,3)}] (-6,0);
\draw[decorate, decoration={random steps,segment length=4pt, amplitude=3pt}] (-12,0) to[out angle=30, in angle=180-30, curve through = {(-11,1) (-8,2)}] (-6,0);

%\draw[decorate, decoration={random steps,segment length=4pt, amplitude=3pt}] (-6,0) to[out angle=180-30, in angle=30, curve through = {(-6,6) (-5,7) (5,6)}] (6,0);
%\draw[decorate, decoration={random steps,segment length=4pt, amplitude=3pt}] (-6,0) to[out angle=30, in angle=180-30, curve through = {(-5,3) (4,2)}] (6,0);

%\draw[decorate, decoration={random steps,segment length=4pt, amplitude=3pt}] (-6,0) to[out angle=180-30, in angle=30, curve through = {(-6,6) (-2,9) (2,7)}] (6,0);

%\draw[decorate, decoration={random steps,segment length=4pt, amplitude=3pt}] (-6,0) to[out angle=180-30, in angle=30, curve through = {(-4,4) (-2,6) (2,4)}] (6,0);

\filldraw[red] (-3,0) circle (2pt) node[black, anchor=north]{$\phi_2(x_3)$};
\filldraw[red] (3,0) circle (2pt) node[black, anchor=north]{$\phi_2(x_4)$};

\filldraw[red] (-12,0) circle (2pt) node[black, anchor=north]{$\phi_2(x_1)$};
\filldraw[red] (-6,0) circle (2pt) node[black, anchor=north]{$\phi_2(x_2)$};

\end{tikzpicture}
    \caption{The two pairs of boundary 2-leg operators insert two non-intersecting SLE bubbles on the upper half-plane anchored to the real axis.}
    \label{fig:2L-2L-4}
\end{figure}
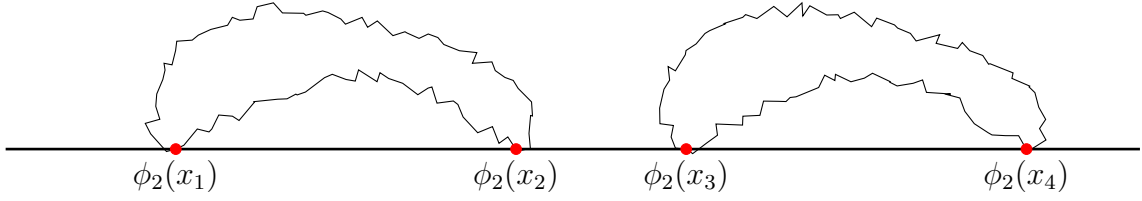

We continue with the four-point function $\braket{ \phi_2(x_1)\phi_2(x_2)\phi_2(x_3)\phi_2(x_4) }$, which describes two SLE bubbles that connect the points $x_1$, $x_2$, $x_3$, and $x_4$, as shown in Figure \ref{fig:2L-2L-4}. Similarly to the previous case, we have
\begin{equation}
    \Braket{ \phi_2(x_1)\phi_2(x_2)\phi_2(x_3)\phi_2(x_4) } = \frac{H(\xi )}{
    (x_1 - x_2)^{2\Delta_{(3,1)}}
    (x_3 - x_4)^{2\Delta_{(3,1)}}
    } \ ,
\end{equation}
where the function $H(\xi)$ solves the third-order BPZ equation. From Figure~\ref{fig:2L-2L-4}, we see that, in the limit $x_2,x_3 \rightarrow x$, or equivalently $\xi \rightarrow 1$, the two SLE bubbles should merge at a pivotal point located on the boundary (as in Figure~\ref{fig:4L_4L_SLE}, but with $z \rightarrow x$). Therefore, we require that
\begin{equation} \label{symx1}
    H(\xi) \overset{\xi \rightarrow 1}\propto (1- \xi)^{\Delta_{(5, 1)}} + \ldots \ .
\end{equation}
Similarly to the case of $\braket{ \phi_1(x_1)\phi_1(x_2)\phi_1(x_3)\phi_1(x_4) }$, the right combination of $h_1(\xi)$, $h_2(\xi)$, and $h_3(\xi)$ in \eqref{h123} can then be obtained by considering the crossing transformation between the solutions in \eqref{h123} and the solutions to \eqref{bpz3rd} with singular point at $\xi = 1$. The explicit expression of this crossing transformation can be found in \cite{nr21}. We find that the right combination of  $h_1(\xi)$, $h_2(\xi)$, and $h_3(\xi)$ is
\begin{align}
    H(\xi)&= h_1(\xi) + c_2 h_2(\xi) + c_3 h_3(\xi) \label{right2} \\
    &= d_1 h^{(1)}_3(\xi) \ , \label{right3}
\end{align}
where $h^{(1)}_3(\xi)$ is the solution with singular point at $\xi = 1$ such that $h^{(1)}_3(\xi) \overset{\xi \rightarrow 1}\propto (1-\xi)^{\Delta_{(5,1)}}$. Moreover, we have
\begin{subequations}
\begin{align}
c_2 &=
\frac{\pi  \left(\csc \left(\frac{8 \pi }{\kappa }\right)+\csc \left(\frac{16 \pi
   }{\kappa }\right)\right) \sec \left(\frac{8 \pi }{\kappa }\right) \Gamma
   \left(2-\frac{8}{\kappa }\right) \Gamma \left(\frac{4}{\kappa }\right) \Gamma
   \left(\frac{\kappa -8}{\kappa }\right) \Gamma \left(\frac{2 (\kappa -6)}{\kappa
   }\right)}{2 \Gamma \left(\frac{8}{\kappa }\right)^2 \Gamma \left(\frac{2 (\kappa
   -8)}{\kappa }\right)^2 \Gamma \left(\frac{\kappa -4}{\kappa }\right)^2}
\\
c_3 &=
-\frac{(\kappa -8) \sec \left(\frac{8 \pi }{\kappa }\right) \Gamma
   \left(\frac{20}{\kappa }-1\right) \Gamma \left(\frac{2 (\kappa -6)}{\kappa
   }\right)}{2 (\kappa -16) \Gamma \left(\frac{12}{\kappa }\right) \Gamma
   \left(\frac{\kappa -4}{\kappa }\right)}
\\
d_1 &=
\frac{(\kappa -8) \left(2 \cos \left(\frac{8 \pi }{\kappa }\right)+1\right)^2 \Gamma
   \left(\frac{\kappa -12}{\kappa }\right) \Gamma \left(\frac{2 (\kappa -6)}{\kappa
   }\right)}{(\kappa -16) \left(2 \cos \left(\frac{8 \pi }{\kappa }\right)+2 \cos
   \left(\frac{16 \pi }{\kappa }\right)+1\right) \Gamma \left(2-\frac{20}{\kappa
   }\right) \Gamma \left(\frac{\kappa -4}{\kappa }\right)}
\ .
\end{align}
\end{subequations}
It is straightforward to check that \eqref{right2} obeys \eqref{symx1}. From \eqref{h123}, we deduce that the structure constant $C_{2, 2 ;2}$ is $\sqrt{-c_2}$, and we conjecture that $c_{2, 2 ;4} = \sqrt{d_1}$ (see \eqref{listC}).

\section{\label{sec:final} Conclusion and outlook}

The emergence of conformal invariance in the scaling limit of lattice critical models, now proved in at least some interesting cases, leads to the assumption that certain aspects of the large scale behavior of these models can be described with the use of conformal field theory (CFT).
At the same time, it is believed and in some cases proved that various interfaces in these models converge to Schramm-Loewner Evolution (SLE) curves.
Consequently, in the scaling limit, it is natural to express certain (renormalized) probabilities and SLE observables as CFT correlation functions that obey consistency conditions, such as appropriate boundary conditions and positivity.

In this pedagogical note, we have identified various probabilities that can be expressed in terms of CFT three-point functions on the upper half-plane involving degenerate fields.
We have then shown how to compute such three-point functions using specific combinations of fundamental solutions of the BPZ equations that are real and non-negative and satisfy appropriate boundary conditions.

We validated this method by re-deriving certain known expressions, such as Schramm's left-passage formula and the SLE Green's function, which have previously been proved rigorously \cite{10.1214/ECP.v6-1041, lawler2008conformally}. Furthermore, we obtained new formulas for various densities of anchored clusters and SLE bubbles. These include, in particular, the density of the outer boundary of a cluster or bubble anchored to two points on the real line and the density of touching points between two separate clusters anchored to the real line.

Percolation, the FK random cluster model, the $O(n)$ loop model, and related loop models are believed to be associated with complex, often logarithmic, CFTs \cite{NR20, rs01, car13, Pearce_2006, cf24, cf24+}.
We have shown that the method discussed in this note has the ability to unify the derivation of various known results and also to generate new interesting predictions. We hope that this ability can help physicists to better understand the CFT structures emerging from loop models, and can inspire mathematicians to prove new results about SLE and related objects.

\bigskip
\noindent{{\bf Acknowledgments.} The first two authors thank Wei Qian for useful conversations at an early stage of the project that led to this article.}

\bibliographystyle{unsrt}
\bibliography{bibliography}

@misc{cf24+,
      title={Conformally covariant probabilities, operator product expansions, ans logarithmic correlations in two-dimensional critical percolation}, 
      author={Camia, Federico and Feng, Yu},
      year={2024},
      eprint={2407.04246},
      archivePrefix={arXiv},
      primaryClass={math-ph},
      url={
https://doi.org/10.48550/arXiv.2407.04246}
}

@article{cf24,
   title={Logarithmic correlation functions in 2D critical percolation},
   volume={2024},
   pages={},
   ISSN={},
   url={},
   DOI={https://doi.org/10.1007/JHEP08(2024)103},
   number={103},
   journal={Journal of High Energy Physics},
   publisher={Springer},
   author={Camia, Federico and Feng, Yu},
   year={2024},
   month=aug}

@article{cgn15,
   title={Planar Ising Magnetization I. Uniqueness of the Critical Scaling Limit},
   volume={43},
   pages={528-571},
   ISSN={},
   url={http://www.jstor.org/stable/24519153},
   DOI={},
   number={2},
   journal={Annals of Probability},
   publisher={Institute of Mathematical Statistics},
   author={Camia, Federico and Garban, Christoph and Newman, Charles M.},
   year={2015}
}

@article{cjn20,
   title={Exponential Decay for the Near-Critical Scaling Limit of the Planar Ising Model},
   volume={73},
   pages={1371-1405},
   ISSN={},
   url={https://doi.org/10.1002/cpa.21884},
   DOI={},
   number={},
   journal={Communications on Pure and Applied Mathematics},
   publisher={Wiley},
   author={Camia, Federico and Jiang, Jianping and Newman, Charles M.},
   year={2020}
}

@article{CamiaApr24,
   title={On the denisty of 2D critical percolation gaskets and anchored clusters},
   volume={114},
   pages={1-10},
   ISSN={0377-9017},
   url={},
   DOI={},
   number={2},
   journal={Letters in Mathematical Physics},
   publisher={Springer},
   author={Camia, Federico},
   year={2024},
   month=apr }

@misc{watts24,
      title={Explicit expressions for Virasoro singular vectors}, 
      author={Gérard M T Watts},
      year={2024},
      eprint={2412.07505},
      archivePrefix={arXiv},
      primaryClass={hep-th},
      url={https://arxiv.org/abs/2412.07505}, 
}

@article{skz07,
   title={Exact factorization of correlation functions in two-dimensional critical percolation},
   volume={76},
   ISSN={1550-2376},
   url={http://dx.doi.org/10.1103/PhysRevE.76.041106},
   DOI={10.1103/physreve.76.041106},
   number={4},
   journal={Physical Review E},
   publisher={American Physical Society (APS)},
   author={Simmons, Jacob J. H. and Kleban, Peter and Ziff, Robert M.},
   year={2007},
   month=oct }

@article{nr21,
   title={Analytic conformal bootstrap and Virasoro primary fields in the  Ashkin-Teller model},
   volume={11},
   ISSN={2542-4653},
   url={http://dx.doi.org/10.21468/SciPostPhys.11.5.089},
   DOI={10.21468/scipostphys.11.5.089},
   number={5},
   journal={SciPost Physics},
   publisher={Stichting SciPost},
   author={Nemkov, Nikita and Ribault, Sylvain},
   year={2021},
   month=nov }

@article{Gamsa_2005,
   title={The scaling limit of two cluster boundaries in critical lattice models},
   volume={2005},
   ISSN={1742-5468},
   url={http://dx.doi.org/10.1088/1742-5468/2005/12/P12009},
   DOI={10.1088/1742-5468/2005/12/p12009},
   number={12},
   journal={Journal of Statistical Mechanics: Theory and Experiment},
   publisher={IOP Publishing},
   author={Gamsa, Adam and Cardy, John},
   year={2005},
   month=dec, pages={P12009–P12009} }

@article{gamsa2006correlation,
  title={Correlation functions of twist operators applied to single self-avoiding loops},
  author={Gamsa, Adam and Cardy, John},
  journal={Journal of Physics A: Mathematical and General},
  volume={39},
  number={41},
  pages={12983--13003},
  year={2006}
}

@article{lgr15,
   title={Minkowski content and natural parameterization for the Schramm-Loewner evolution},
   volume={43},
   ISSN={0091-1798},
   url={http://dx.doi.org/10.1214/13-AOP874},
   DOI={10.1214/13-aop874},
   number={3},
   journal={The Annals of Probability},
   publisher={Institute of Mathematical Statistics},
   author={Lawler, Gregory F. and Rezaei, Mohammad A.},
   year={2015},
   month=may }

@article{Cardy_2006,
   title={The O(n) Model on the Annulus},
   volume={125},
   ISSN={1572-9613},
   url={http://dx.doi.org/10.1007/s10955-006-9186-8},
   DOI={10.1007/s10955-006-9186-8},
   number={1},
   journal={Journal of Statistical Physics},
   publisher={Springer Science and Business Media LLC},
   author={Cardy, John},
   year={2006},
   month=aug, pages={1-21} }

@misc{smirnov07,
      title={Conformal invariance in random cluster models. I. Holomorphic fermions in the Ising model}, 
      author={Stanislav Smirnov},
      year={2007},
      eprint={0708.0039},
      archivePrefix={arXiv},
      primaryClass={math-ph},
      url={https://arxiv.org/abs/0708.0039}, 
}

@article{Bauer2006,
   title={2D growth processes: SLE and Loewner chains},
   volume={432},
   ISSN={0370-1573},
   url={http://dx.doi.org/10.1016/j.physrep.2006.06.002},
   DOI={10.1016/j.physrep.2006.06.002},
   number={3?4},
   journal={Physics Reports},
   publisher={Elsevier BV},
   author={Bauer, Michel and Bernard, Denis},
   year={2006},
   month=oct, pages={115-221} }

@article{Bauer2002,
   title={SLE$_\kappa$ growth processes and conformal field theories},
   volume={543},
   ISSN={0370-2693},
   url={http://dx.doi.org/10.1016/S0370-2693(02)02423-1},
   DOI={10.1016/s0370-2693(02)02423-1},
   number={1?2},
   journal={Physics Letters B},
   publisher={Elsevier BV},
   author={Bauer, Michel and Bernard, Denis},
   year={2002},
   month=sep, pages={135-138} }

@article{PhysRevLett.88.130601,
  title = {Monte Carlo Tests of Stochastic Loewner Evolution Predictions for the 2D Self-Avoiding Walk},
  author = {Kennedy, Tom},
  journal = {Phys. Rev. Lett.},
  volume = {88},
  issue = {13},
  pages = {130601},
  numpages = {4},
  year = {2002},
  month = {Mar},
  publisher = {American Physical Society},
  doi = {10.1103/PhysRevLett.88.130601},
  url = {https://link.aps.org/doi/10.1103/PhysRevLett.88.130601}
}

@misc{lawler2002,
      title={On the scaling limit of planar self-avoiding walk}, 
      author={Gregory F. Lawler and Oded Schramm and Wendelin Werner},
      year={2002},
      eprint={math/0204277},
      archivePrefix={arXiv},
      primaryClass={math.PR},
      url={https://arxiv.org/abs/math/0204277}, 
}

@misc{cn06,
      title={Critical Percolation Exploration Path and SLE(6): a Proof of Convergence}, 
      author={Federico Camia and Charles M. Newman},
      year={2006},
      eprint={math/0604487},
      archivePrefix={arXiv},
      primaryClass={math.PR},
      url={https://arxiv.org/abs/math/0604487}, 
}

@misc{schramm99,
      title={Scaling limits of loop-erased random walks and uniform spanning trees}, 
      author={Oded Schramm},
      year={1999},
      eprint={math/9904022},
      archivePrefix={arXiv},
      primaryClass={math.PR},
      url={https://arxiv.org/abs/math/9904022}, 
}

@article{Copin2021,
   title={Planar random-cluster model: fractal properties of the critical phase},
   volume={181},
   ISSN={1432-2064},
   url={http://dx.doi.org/10.1007/s00440-021-01060-6},
   DOI={10.1007/s00440-021-01060-6},
   number={1?3},
   journal={Probability Theory and Related Fields},
   publisher={Springer Science and Business Media LLC},
   author={Duminil-Copin, Hugo and Manolescu, Ioan and Tassion, Vincent},
   year={2021},
   month=jun, pages={401-449} }

@article{smirnov01,
title = {Critical percolation in the plane: conformal invariance, Cardy's formula, scaling limits},
journal = {Comptes Rendus de l'AcadÃ©mie des Sciences - Series I - Mathematics},
volume = {333},
number = {3},
pages = {239-244},
year = {2001},
issn = {0764-4442},
doi = {https://doi.org/10.1016/S0764-4442(01)01991-7},
url = {https://www.sciencedirect.com/science/article/pii/S0764444201019917},
author = {Stanislav Smirnov}
}

@article{Cardy92,
   title={Critical percolation in finite geometries},
   volume={25},
   ISSN={1361-6447},
   url={http://dx.doi.org/10.1088/0305-4470/25/4/009},
   DOI={10.1088/0305-4470/25/4/009},
   number={4},
   journal={Journal of Physics A: Mathematical and General},
   publisher={IOP Publishing},
   author={Cardy, J L},
   year={1992},
   month=feb, pages={L201-L206} }

@article{bkw76,
doi = {10.1088/0305-4470/9/3/009},
url = {https://dx.doi.org/10.1088/0305-4470/9/3/009},
year = {1976},
month = {mar},
publisher = {},
volume = {9},
number = {3},
pages = {397},
author = {R J Baxter and  S B Kelland and  F Y Wu},
title = {Equivalence of the Potts model or Whitney polynomial with an ice-type model},
journal = {Journal of Physics A: Mathematical and General},
}

@article{gnjrs23,
    author = "Grans-Samuelsson, Linnea and Jacobsen, Jesper Lykke and Nivesvivat, Rongvoram and Ribault, Sylvain and Saleur, Hubert",
    title = "{From combinatorial maps to correlation functions in loop models}",
    eprint = "2302.08168",
    archivePrefix = "arXiv",
    primaryClass = "hep-th",
    doi = "10.21468/SciPostPhys.15.4.147",
    journal = "SciPost Phys.",
    volume = "15",
    number = "4",
    pages = "147",
    year = "2023"
}

@article{ksz06,
    author = "Kleban, P. and Simmons, J. J. H. and Ziff, R. M.",
    title = "{Anchored Critical Percolation Clusters and 2-D Electrostatics}",
    eprint = "cond-mat/0605120",
    archivePrefix = "arXiv",
    doi = "10.1103/PhysRevLett.97.115702",
    journal = "Phys. Rev. Lett.",
    volume = "97",
    pages = "115702",
    year = "2006"
}

@article{Cardy84,
    author = "Cardy, John L.",
    title = "{Conformal Invariance and Surface Critical Behavior}",
    doi = "10.1016/0550-3213(84)90241-4",
    journal = "Nucl. Phys. B",
    volume = "240",
    pages = "514--532",
    year = "1984"
}

@article{Cardy_2005,
   title={SLE for theoretical physicists},
   volume={318},
   ISSN={0003-4916},
   url={http://dx.doi.org/10.1016/j.aop.2005.04.001},
   DOI={10.1016/j.aop.2005.04.001},
   number={1},
   journal={Annals of Physics},
   publisher={Elsevier BV},
   author={Cardy, John},
   year={2005},
   month=jul, pages={81-118} }

@article{Nienhuis82,
	author = {Nienhuis, Bernard},
	doi = {10.1103/PhysRevLett.49.1062},
	issue = {15},
	journal = {Phys. Rev. Lett.},
	month = {Oct},
	numpages = {0},
	pages = {1062--1065},
	publisher = {American Physical Society},
	title = {Exact Critical Point and Critical Exponents of $\mathrm{O}(n)$ Models in Two Dimensions},
	url = {https://link.aps.org/doi/10.1103/PhysRevLett.49.1062},
	volume = {49},
	year = {1982}}

@article{DF84,
	author = {Vl.S. Dotsenko and V.A. Fateev},
	doi = {https://doi.org/10.1016/0550-3213(84)90269-4},
	issn = {0550-3213},
	journal = {Nuclear Physics B},
	number = {3},
	pages = {312-348},
	title = {Conformal algebra and multipoint correlation functions in 2D statistical models},
	url = {https://www.sciencedirect.com/science/article/pii/0550321384902694},
	volume = {240},
	year = {1984}}

@article{NR20,
	archiveprefix = {arXiv},
	author = {Nivesvivat, Rongvoram and Ribault, Sylvain},
	doi = {10.21468/SciPostPhys.10.1.021},
	eprint = {2007.04190},
	journal = {SciPost Phys.},
	number = {1},
	pages = {021},
	primaryclass = {hep-th},
	title = {{Logarithmic CFT at generic central charge: from Liouville theory to the $Q$-state Potts model}},
	volume = {10},
	year = {2021}}

@article{mr07,
	archiveprefix = {arXiv},
	author = {Mathieu, Pierre and Ridout, David},
	doi = {10.1016/j.nuclphysb.2008.02.017},
	eprint = {0711.3541},
	journal = {Nucl. Phys. B},
	pages = {268--295},
	primaryclass = {hep-th},
	title = {{Logarithmic M(2,p) minimal models, their logarithmic couplings, and duality}},
	volume = {801},
	year = {2008}}

@article{fk72,
	author = {C.M. Fortuin and P.W. Kasteleyn},
	doi = {10.1016/0031-8914(72)90045-6},
	issn = {0031-8914},
	journal = {Physica},
	number = {4},
	pages = {536 - 564},
	title = {On the random-cluster model},
	volume = {57},
	year = {1972}}

@article{car13,
	archiveprefix = {arXiv},
	author = {Cardy, John},
	doi = {10.1088/1751-8113/46/49/494001},
	eprint = {1302.4279},
	journal = {J. Phys.},
	pages = {494001},
	primaryclass = {cond-mat.stat-mech},
	slaccitation = {%%CITATION = ARXIV:1302.4279;%%},
	title = {{Logarithmic conformal field theories as limits of ordinary CFTs and some physical applications}},
	volume = {A46},
	year = {2013}}

@article{fsz87,
	author = {Di Francesco, Philippe and Saleur, Hubert and Zuber, Jean-Bernard},
	journal = {Journal of statistical physics},
	number = {1-2},
	pages = {57--79},
	publisher = {Springer},
	title = {Relations between the Coulomb gas picture and conformal invariance of two-dimensional critical models},
	volume = {49},
	year = {1987}}

@article{rs01,
	archiveprefix = {arXiv},
	author = {Read, N. and Saleur, H.},
	doi = {10.1016/S0550-3213(01)00395-9},
	eprint = {hep-th/0106124},
	journal = {Nucl. Phys.},
	pages = {409},
	primaryclass = {hep-th},
	slaccitation = {%%CITATION = HEP-TH/0106124;%%},
	title = {{Exact spectra of conformal supersymmetric nonlinear sigma models in two-dimensions}},
	volume = {B613},
	year = {2001}}

@article{zam84,
	author = {Zamolodchikov, Al.B.},
	doi = {10.1007/BF01214585},
	journal = {Commun.Math.Phys.},
	pages = {419-422},
	title = {{Conformal symmetry in two dimensions: an explicit recurrence formula for the conformal partial wave amplitude}},
	volume = {96},
	year = {1984}}

@book{fms97,
	author = {Di Francesco, P. and Mathieu, P. and S{\'e}n{\'e}chal, D.},
	doi = {10.1007/978-1-4612-2256-9},
	title = {Conformal field theory},
	year = {1997},
    publisher = {Springer-Verlag New York, Inc.}}

@article{bpz84,
	author = {Belavin, A. A. and Polyakov, Alexander M. and Zamolodchikov, A. B.},
	doi = {10.1016/0550-3213(84)90052-X},
	journal = {Nucl. Phys.},
	note = {[,605(1984)]},
	pages = {333-380},
	reportnumber = {CERN-TH-3827},
	slaccitation = {%%CITATION = NUPHA,B241,333;%%},
	title = {{Infinite Conformal Symmetry in Two-Dimensional Quantum Field Theory}},
	volume = {B241},
	year = {1984}}

@article{fzz00,
  title={Boundary Liouville field theory I. Boundary state and boundary two-point function},
  author={Fateev, V and Zamolodchikov, Alexander and Zamolodchikov, Al},
  journal={arXiv preprint hep-th/0001012},
  year={2000}
}

@article{lew92,
	author = {Lewellen, David C.},
	journal = {Nucl. Phys.},
	pages = {654-682},
	slaccitation = {%%CITATION = NUPHA,B372,654;%%},
	title = {Sewing constraints for conformal field theories on surfaces with boundaries},
	volume = {B372},
	year = {1992}}

@article{Kadanoff_1978,
doi = {10.1088/0305-4470/11/7/027},
url = {https://doi.org/10.1088/0305-4470/11/7/027},
year = {1978},
month = {jul},
publisher = {},
volume = {11},
number = {7},
pages = {1399},
author = {L P Kadanoff},
title = {Lattice Coulomb gas representations of two-dimensional problems},
journal = {Journal of Physics A: Mathematical and General},
}

@article{nienhuis1984critical,
  title={Critical behavior of two-dimensional spin models and charge asymmetry in the Coulomb gas},
  author={Nienhuis, Bernard},
  journal={Journal of Statistical Physics},
  volume={34},
  number={5},
  pages={731--761},
  year={1984},
  publisher={Springer}
}

@article{DUPLANTIER1989229,
title = {Two-dimensional fractal geometry, critical phenomena and conformal invariance},
journal = {Physics Reports},
volume = {184},
number = {2},
pages = {229-257},
year = {1989},
issn = {0370-1573},
doi = {https://doi.org/10.1016/0370-1573(89)90042-2},
url = {https://www.sciencedirect.com/science/article/pii/0370157389900422},
author = {Bertrand Duplantier}
}

@Inbook{Jacobsen2009,
author="Jacobsen, Jesper Lykke",
title="Conformal Field Theory Applied to Loop Models",
bookTitle="Polygons, Polyominoes and Polycubes",
year="2009",
publisher="Springer Netherlands",
address="Dordrecht",
pages="347--424",
isbn="978-1-4020-9927-4",
doi="10.1007/978-1-4020-9927-4_14",
url="https://doi.org/10.1007/978-1-4020-9927-4_14"
}

@article{kager2004guide,
  title={A guide to stochastic L{\"o}wner evolution and its applications},
  author={Kager, Wouter and Nienhuis, Bernard},
  journal={Journal of statistical physics},
  volume={115},
  number={5},
  pages={1149--1229},
  year={2004},
  publisher={Springer}
}

@book{lawler2008conformally,
  title={Conformally invariant processes in the plane},
  author={Lawler, Gregory F},
  volume={114},
  year={2008},
  publisher={American Mathematical Soc.}
}

@Incollection{tsirelson2004lectures,
  title={Random Planar Curves and Schramm-Loewner Evolutions},
  booktitle={Lectures on Probability Theory and Statistics: Ecole D'Et{\'e} de Probabilit{\'e}s de Saint-Flour XXXII-2002},
  author={Werner, Wendelin},
  year={2004},
  publisher={Springer}
}

@article{CRMATH_2014__352_2_157_0,
     author = {Dmitry Chelkak and Hugo Duminil-Copin and Cl\'ement Hongler and Antti Kemppainen and Stanislav Smirnov},
     title = {Convergence of {Ising} interfaces to {Schramm's} {SLE} curves},
     journal = {Comptes Rendus. Math\'ematique},
     pages = {157--161},
     year = {2014},
     publisher = {Elsevier},
     volume = {352},
     number = {2},
     doi = {10.1016/j.crma.2013.12.002},
     language = {en},
}

@article{10.4310/MRL.2001.v8.n6.a4,
author = {Stanislav Smirnov and Wendelin Werner},
title = {{Critical exponents for two-dimensional percolation}},
volume = {8},
journal = {Mathematical Research Letters},
number = {6},
publisher = {International Press of Boston},
pages = {729 -- 744},
year = {2001},
doi = {10.4310/MRL.2001.v8.n6.a4},
URL = {https://doi.org/10.4310/MRL.2001.v8.n6.a4}
}

@article{zbMATH06446409,
 author = {Chelkak, Dmitry and Hongler, Cl{\'e}ment and Izyurov, Konstantin},
 title = {Conformal invariance of spin correlations in the planar {Ising} model},
 fjournal = {Annals of Mathematics. Second Series},
 journal = {Ann. Math. (2)},
 issn = {0003-486X},
 volume = {181},
 number = {3},
 pages = {1087--1138},
 year = {2015},
 language = {English},
 doi = {10.4007/annals.2015.181.3.5},
 zbMATH = {6446409},
 Zbl = {1318.82006}
}

@article{dubedat2006euler,
  title={Euler integrals for commuting SLEs},
  author={Dub{\'e}dat, Julien},
  journal={Journal of statistical physics},
  volume={123},
  number={6},
  pages={1183--1218},
  year={2006},
  publisher={Springer}
}

@article{10.1214/07-AOP364,
author = {Vincent Beffara},
title = {{The dimension of the SLE curves}},
volume = {36},
journal = {The Annals of Probability},
number = {4},
publisher = {Institute of Mathematical Statistics},
pages = {1421 -- 1452},
year = {2008},
doi = {10.1214/07-AOP364},
URL = {https://doi.org/10.1214/07-AOP364}
}

@article{GURARIE1993535,
title = {Logarithmic operators in conformal field theory},
journal = {Nuclear Physics B},
volume = {410},
number = {3},
pages = {535-549},
year = {1993},
issn = {0550-3213},
doi = {https://doi.org/10.1016/0550-3213(93)90528-W},
url = {https://www.sciencedirect.com/science/article/pii/055032139390528W},
author = {V. Gurarie}
}

@article{Vasseur_2012,
doi = {10.1088/1742-5468/2012/07/L07001},
url = {https://doi.org/10.1088/1742-5468/2012/07/L07001},
year = {2012},
month = {jul},
publisher = {IOP Publishing and SISSA},
volume = {2012},
number = {07},
pages = {L07001},
author = {Vasseur, Romain and Jacobsen, Jesper Lykke and Saleur, Hubert},
title = {Logarithmic observables in critical percolation},
journal = {Journal of Statistical Mechanics: Theory and Experiment}
}

@article{10.1214/ECP.v6-1041,
author = {Oded Schramm},
title = {{A Percolation Formula}},
volume = {6},
journal = {Electronic Communications in Probability},
number = {none},
publisher = {Institute of Mathematical Statistics and Bernoulli Society},
pages = {115 -- 120},
year = {2001},
doi = {10.1214/ECP.v6-1041},
URL = {https://doi.org/10.1214/ECP.v6-1041}
}

@article{Pearce_2006,
doi = {10.1088/1742-5468/2006/11/P11017},
url = {https://doi.org/10.1088/1742-5468/2006/11/P11017},
year = {2006},
month = {nov},
publisher = {},
volume = {2006},
number = {11},
pages = {P11017},
author = {Pearce, Paul A and Rasmussen, Jørgen and Zuber, Jean-Bernard},
title = {Logarithmic minimal models},
journal = {Journal of Statistical Mechanics: Theory and Experiment}
}

@article{chelkak2012universality,
  title={Universality in the 2D Ising model and conformal invariance of fermionic observables},
  author={Chelkak, Dmitry and Smirnov, Stanislav},
  journal={Inventiones mathematicae},
  volume={189},
  number={3},
  pages={515--580},
  year={2012},
  publisher={Springer}
}

@article{10.1214/EJP.v3-32,
author = {Harry Kesten and Vladas Sidoravicius and Yu Zhang},
title = {{Almost All Words Are Seen In Critical Site Percolation On The Triangular Lattice}},
volume = {3},
journal = {Electronic Journal of Probability},
number = {none},
publisher = {Institute of Mathematical Statistics and Bernoulli Society},
pages = {1 -- 75},
year = {1998},
doi = {10.1214/EJP.v3-32},
URL = {https://doi.org/10.1214/EJP.v3-32}
}

@article{10.1214/16-EJP3452,
author = {Dmitry Chelkak and Hugo Duminil-Copin and Cl{\'e}ment Hongler},
title = {{Crossing probabilities in topological rectangles for the critical planar FK-Ising model}},
volume = {21},
journal = {Electronic Journal of Probability},
number = {none},
publisher = {Institute of Mathematical Statistics and Bernoulli Society},
pages = {1 -- 28},
year = {2016},
doi = {10.1214/16-EJP3452},
URL = {https://doi.org/10.1214/16-EJP3452}
}

@article{PhysRevLett.83.1359,
  title = {Path-Crossing Exponents and the External Perimeter in 2D Percolation},
  author = {Aizenman, Michael and Duplantier, Bertrand and Aharony, Amnon},
  journal = {Phys. Rev. Lett.},
  volume = {83},
  issue = {7},
  pages = {1359--1362},
  numpages = {0},
  year = {1999},
  month = {Aug},
  publisher = {American Physical Society},
  doi = {10.1103/PhysRevLett.83.1359},
  url = {https://link.aps.org/doi/10.1103/PhysRevLett.83.1359}
}

@article{laslier2024tilted,
  title={Tilted Solid-On-Solid is liquid: scaling limit of SOS with a potential on a slope},
  author={Laslier, Beno{\^\i}t and Lubetzky, Eyal},
  journal={arXiv preprint arXiv:2409.08745},
  year={2024}
}

\end{document}